\documentclass[useAMS,usenatbib]{mnras}

\usepackage{graphicx}
\usepackage{lscape}

\title[Catalogue and Properties of $\delta$~Sct Stars in Binaries]{Catalogue and Properties of $\delta$~Scuti Stars in Binaries}
\author[A. Liakos and P. Niarchos]{Alexios Liakos$^{1}$\thanks{E-mail: alliakos@noa.gr} and
Panagiotis Niarchos$^{2}$\\
$^{1}$Institute for Astronomy, Astrophysics, Space Applications and Remote Sensing, National Observatory of Athens,\\ Metaxa \& Vas. Pavlou St., GR-15236, Penteli, Athens, Hellas\\
$^{2}$Department of Astrophysics, Astronomy and Mechanics, Faculty of Physics, National \& Kapodistrian University of Athens, \\ Univ. Campus, GR-15784, Zografos, Athens, Hellas\\}

\begin{document}

\date{Accepted 201X December XX. Received 201X December XX; in original form 201X October XX}

\pagerange{\pageref{firstpage}--\pageref{lastpage}} \pubyear{201X}

\maketitle

\label{firstpage}

\begin{abstract}
The catalogue contains 199 confirmed cases of binary systems containing at least one pulsating component of $\delta$~Sct type. The sample is divided into subgroups in order to describe the properties and characteristics of the $\delta$~Sct type stars in binaries according to their Roche geometry. Demographics describing quantitatively our knowledge for these systems as well as the distributions of their pulsating components in the Mass-Radius, Colour-Magnitude and Evolutionary Status-Temperature diagrams are presented and discussed. It is shown that a threshold of $\sim$13~days of the orbital period regarding the influence of binarity on the pulsations is established. Finally, the correlations between the pulsation periods and the orbital periods, evolutionary status, and companion's gravity influence are updated based on the largest sample to date.

\end{abstract}


\section{Introduction}
\label{sec:s1}
The single pulsating stars of $\delta$~Scuti type range between A-F spectral types and have mass values between 1.4-3~M$_{\sun}$. They lie inside the classical instability strip and their luminosity classes range between III-V \citep{BRE00}. They potentially exhibit radial and non radial pulsations with frequencies ranging between 3-80~cycle~d$^{-1}$ \citep{BRE00} and can oscillate in both $p$ and $g$ modes. The origin and the evolution of their intrinsic oscillations are in general explained by the $\kappa$-mechanism mostly due to partial ionization of HeII \citep[][and references therein]{HOU99, BAL15} and with a small contribution of the combined HI and HeI ionization zone \citep{CAS68}. However, recent results by \citet{ANT14} and \citet{GRA15} show that the $\kappa$-mechanism is insufficient to explain all the frequency modes, especially the high-order, non-radial ones. According to these works the turbulent pressure in the Hydrogen convective zone can be considered as an additional pulsations driving mechanism.

Binary systems are stellar objects that can be considered as the most powerful tools for astronomers to derive absolute stellar parameters (e.g. masses, radii, luminosities) and to create stellar evolutionary models. The eclipsing systems, in particular, provide the means for accurate determination of those parameters, which can be easily calculated by modelling their light and radial velocity curves (hereafter RVs). Moreover, binary systems also provide an extra tool, namely the orbital period analysis (i.e. $O-C$ or $ETV$ analysis), which is essential for understanding the binary interaction and the physical mechanisms (i.e. mass transfer) that modulate their orbital period.

$\delta$~Scuti stars in binary systems show slight difference in comparison with the single ones regarding the evolutionary stage, since they are located mostly inside the Main Sequence band \citep{LIA12} and only few of them have left it. On the other hand, single and binary $\delta$~Scuti stars show similar oscillation properties. The first notification of this kind of systems was made by \citet{MKR02}, who introduced the Algol-type mass accreting pulsators of (B)A-F type as a new class of stellar systems that is commonly referred as `$oEA~stars$' (i.e. oscillating eclipsing systems of Algol type). It should be noticed that this definition is general and can also be applied to binaries that include pulsating stars of another type (e.g. $\gamma$~Doradus, $\beta$~Cephei, RR~Lyrae, Classical Cepheids), but until 2002 systematic works had been made only for binaries with $\delta$~Scuti stars. According to \citet{MKR03} and \citet{SOY366} the discrepancy in evolution between single and binary members $\delta$~Scuti stars can be explained by the mass transfer and the tidal distortions caused by the companion of the pulsator during or even before its Main Sequence life.

The history of the pulsating stars in binaries covers the last $\sim$25 years, while, in particular for the binary $\delta$~Scuti stars, during the last $\sim$10 years a lot of research papers have been published showing significant evolution in our understanding about their properties. In general, the systematic cataloguing of binary pulsators began from \citet{SZA90}, who published a catalogue with all the known pulsating stars that potentially belong to binary systems. Ten years later, due to the increasing number of discoveries, a need for further classification according to the pulsations type was revealed. \citet{ROD00} published a catalogue with all the known $\delta$~Sct stars and pointed out that 86 of them belong to binary or multiple systems. \citet{ROD01}, based on the catalogue of \citet{ROD00}, listed tables with 15 $\delta$~Sct stars in total that are members of binaries. The first results concerning physical properties of the binary $\delta$~Sct stars were published by \citet{SOY366}, who noticed for the first time the connection between pulsation and orbital periods for eclipsing binaries with a $\delta$~Sct member using a sample of 20 systems. A few months later, \citet{SOY370} published lists with confirmed and candidate such systems. A few years later, \citet{ZHO10} published a catalogue containing 89 systems with an oscillating member and distinguished them according to their pulsational properties. \citet{LIA12}, after a long term observational campaign on more than 100 eclipsing systems-candidates for including a $\delta$~Sct component, published an updated catalogue with 75 confirmed cases and derived new correlations between their fundamental stellar characteristics. \citet{ZHA13} presented the first theoretical approach for the connection between orbital and dominant pulsation periods for specifically $\delta$~Sct stars in binaries. \citet{CHA13} published the most recent catalogue of $\delta$~Sct stars (1578 in total) including 141 cases that are members of binary or multiple systems. Finally, \citet{LIA15, LIA16} published an updated catalogue containing the currently known binary $\delta$~Sct stars and noticed for first time that there is a threshold in the orbital period beyond which there is no influence of the binarity to the pulsations.

Similar studies for other pulsating stars in binaries were made by \citet{CAK16}, who found similar correlations between pulsation-orbital periods and pulsation period - $\log g$ for the $\gamma$~Dor-binary members, and by \citet{SKA16} who published a review for RR~Lyrae type pulsators in binary systems.

The present catalogue contains all the currently known binaries with at least one $\delta$~Sct member. Each system included here was extensively checked by using all the available information from the literature, was characterized and categorized according to its Roche geometry. The novelty of this catalogue in comparison with the respective previous ones \citep[e.g.][]{SOY370, LIA12, LIA15, LIA16} is that it also contains systems on wide orbits (i.e. visual binaries) as well as systems with unseen companions \citep[i.e. their binarity was detected through dominant pulsation frequency modulations, see][]{SHI12}. Each individual case is given the corresponding literature references that contain exactly the information for its properties (e.g. basic orbital and physical parameters, pulsation analysis). This provides the means for the reader to directly check each value listed as well as, in some cases, the chronicle of the system. In addition, the apparent limit of $\sim$13~days \citep{LIA15, LIA16}, that separates the systems whose intrinsic pulsations of their $\delta$~Sct stars are affected by the binarity, seems to be well established. Given that the present sample is much larger than all those used in previous studies for the correlations between fundamental stellar characteristics of these systems, updated ones are also presented. Finally, one of the most important contributions of this paper is the list with the ambiguous cases, which can be used for future target selection for observations with the aim to check them for pulsations and/or binarity.

\section{The catalogue}
\label{sec:s2}
The catalogue consists of five lists given in Tables \ref{tab:tab1}-\ref{tab:tab5} of this paper. It contains all of the currently known binaries with a $\delta$~Sct component, ambiguous ones, and others that were incorrectly reported as being of this type. This catalogue is online\footnote{\url{http://alexiosliakos.weebly.com/catalogue.html}} and is frequently updated.

Tables~\ref{tab:tab1}, \ref{tab:tab2}, and \ref{tab:tab3} include all of the currently known binary systems (199 systems) hosting at least one pulsating component of $\delta$~Sct type (203 pulsators). The whole sample is divided into two main groups according to the orbital period ($P_{\rm orb}$) value of the systems. Particularly, Table \ref{tab:tab1} includes the 118 systems with $P_{\rm orb}<13$~days, while Table \ref{tab:tab2} those with $P_{\rm orb}>13$~days. This separation of the sample is based on the arguments given by \citet[][i.e. the pulsations are not affected by the binarity in systems with $P_{\rm orb}>13$~d]{LIA15, LIA16} and is further discussed in Section \ref{sec:s3}. The sample of each table is further subdivided according to the Roche geometry of the systems, since the properties of each subgroup are different (see Sections \ref{sec:s4}, \ref{sec:s5}). In particular, they are divided into `Semi-detached' (the non-pulsating star fills its Roche lobe), `Detached', and `Unclassified' systems (i.e. unknown geometrical configuration). Tables~\ref{tab:tab1} and \ref{tab:tab2} include: the name of the system, its orbital period ($P_{\rm orb}$), the dominant pulsation frequency ($f_{\rm dom}$) (i.e. the frequency with the largest amplitude) of its $\delta$~Sct star and its corresponding amplitude ($A$) in a given filter (e.g. $B$, $V$, $b$, $y$, $W$=white light, $Kp$=Kepler, $Hp$=HIPPARCOS, $CoR$=CoRoT), and the system's type of variation ($ToV$) (i.e. $E$=Eclipsing Binary, $SB$=Spectroscopic Binary; 1=Single line; 2=Double line, $EV$=Ellipsoidal Variable, $Vis$=Visual Binary, $O-C$=system with its $\delta$~Sct star showing cyclic variation of its dominant pulsation period). Moreover, the absolute parameters of the $\delta$~Sct star ($M_{\rm pul},~R_{\rm pul},~T_{\rm pul},~\log g_{\rm pul}$) as well as those of its companion ($M_{\rm comp},~R_{\rm comp}$) are also listed. The symbols $M,~R,~T,~\log g$ correspond to the standard quantities i.e. mass, radius, temperature and gravity acceleration value, respectively. Moreover, the orbital separation of the system's components ($a$) and their mass ratio ($q$) are also presented. In many cases some of the values of the absolute parameters are marked with an asterisk (*). In general, many authors, in absence of RVs measurements, assume the mass value of the primary component according to its spectral type and its luminosity class (i.e. MS star). The rest parameters, however, follow this assumption \citep[e.g.][]{SOY09, SOY11, LIA13b, LEE16b, LEE16a, GUO16}. Table~\ref{tab:tab3} contains binaries with a $\delta$~Sct component for which no exact $P_{\rm orb}$ values have been determined so far. These systems are also considered as unclassified ones, but they are not used in any of the following correlations (Section \ref{sec:s4}). In the last column of Tables~\ref{tab:tab1}-\ref{tab:tab3} the respective reference codes for each system are given and they are fully explained at the bottom of the tables. For many systems additional comments (e.g. cross-identification names, special properties) are also given as tablenotes and they are very useful for literature search. The orbital period values for the majority of the eclipsing binaries were taken from \citet{KRE01, KRE04}, while for the $Kepler$ systems from \citet{SLA11} and \citet{KIR16}. In Tables~\ref{tab:tab1}-\ref{tab:tab3} the errors are given alongside values in parentheses and they correspond to the last digit(s).

Table~\ref{tab:tab4} includes the cases which had been previously reported as binaries with a $\delta$~Sct member, but now they are considered as ambiguous ones. The challenging reasons for these cases to be members of the binary $\delta$~Sct stars group are: a) questionable binarity, b) questionable pulsational behaviour, and c) incorrect classification in the first place. The majority of these cases were reported in the catalogues of \citet{SZA90}, \citet{ROD00}, \citet{ZHO10}, and \citet{CHA13}, while the most of the $Kepler$ stars/systems were listed by \citet{UYT11} and \citet{BRA15}. Table~\ref{tab:tab5} contains the rejected cases of binaries with a $\delta$~Sct component. Most of them were also listed in the aforementioned catalogues, but it was found that either they are not binaries or they do not exhibit pulsations of $\delta$~Sct type. Each individual case of these tables was examined carefully and the final comments are based on an extensive literature search. The first column contains the name of the star/system, the second one comments for each case, while in the last one are given the relevant reference codes.

In all tables the objects are presented in alphabetical order according to either their GCVS ($General~Catalogue~of~Variable~Stars$) designation or to their individual catalogue name. In general, the reference codes are written in the form `$XXXNN$', where `$XXX$' corresponds to the first three letters of the first author's surname and `$NN$' to the year of the publication. The reference codes in the last columns of the tables are given in chronological order. At the bottom of the tables, the references are given in alphabetical order. It should be mentioned that only the most important references are given for each case and it is recommended for the reader to also check the references therein for a more comprehensive view.

The results from light curve modelling and pulsation analysis for KIC~06629588 and KIC~06669809 were briefly presented by \citet{LIA16}, while the results (Roche geometry model and pulsation analysis) for KIC~10619109, KIC~11175495, KIC~10686876, and WX~Dra (=KIC~10581918) are presented here for first time. For all these systems a detailed analysis will be presented in a future paper. The detailed results for GQ~Dra are not yet published, but the value of $f_{\rm dom}$ was kindly provided through a private communication with B. Ulas. \citet{GAU14} published preliminary results of the frequency analysis for KIC~04570326, KIC~05872506, KIC~06541245, KIC~11401845, and KIC~11973705 without giving the exact values of $f_{\rm dom}$, but only their frequency range. The systems which have been found to include two pulsators of $\delta$~Sct type (RS~Cha, $\delta$~Del, KIC~09851944, and KIC~10080943) are given two tablelines, one for each pulsating component (referred as A and B). It should to be noted that the light curves of all $Kepler$ binaries mentioned in this study are available online\footnote{\url{http://keplerebs.villanova.edu/}}.

\section{Demographics}
\label{sec:s3}
In this section we present various demographical diagrams showing distributions of the binaries with a $\delta$~Sct component according to their physical and/or geometrical properties (e.g. Mass, radius, Roche geometry).

\begin{figure}
\includegraphics[width=7.6cm]{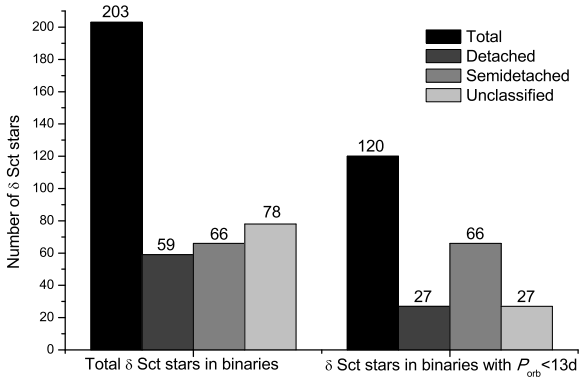}
\caption{Demographics of $\delta$~Sct stars in binaries according to the Roche geometry (detached, semi-detached, unclassified) of the systems. Left columns refer to $\delta$~Sct stars of all systems, while the right ones to those that belong to systems with $P_{\rm orb}<13$~days.}
    \label{fig:stat1}
\end{figure}

\begin{figure}
\includegraphics[width=\columnwidth]{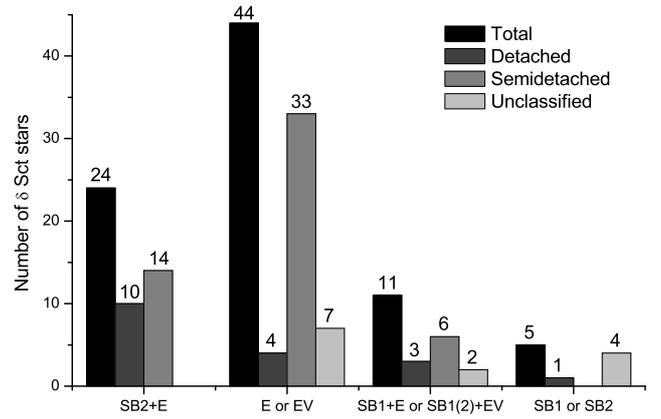}
\caption{Demographics of $\delta$~Sct stars in binaries (with $P_{\rm orb}<13$~days) and with known absolute parameters distributed according to the geometrical type of the systems. The sample is further subdivided according to the available information for the calculation of the absolute parameters. For details see Section~\ref{sec:s3}.}
    \label{fig:stat2}
\end{figure}

In Fig.~\ref{fig:stat1} the demographics of the sample of the binary $\delta$~Sct stars according to the Roche geometry configuration of the system in which they belong is presented. In the same plot, similar demographics are also given for the binary $\delta$~Sct stars in systems with $P_{\rm orb}<13$~days, since they are the only ones that are used for the correlations of physical parameters in the next section. In Fig.~\ref{fig:stat2} the demographics of the binary $\delta$~Sct stars in systems with $P_{\rm orb}<13$~days and with known absolute parameters is also presented. In this graph, the sample is divided into three different groups according to the information used for the calculation of their absolute parameters. In particular, the first group (SB2+E) contains the $\delta$~Sct stars in systems which are both eclipsing and double-line spectroscopic binaries and whose their absolute parameters were calculated from data derived from the analysis of RVs and light curves. The second group (E or EV) includes the $\delta$~Sct stars in eclipsing and ellipsoidal systems for which there is no spectroscopic information so far. The calculation of their absolute parameters is based only on the light curves solution and on assumptions for the mass of the primary star according to theoretical models. The third group (SB1+E or SB1(2)+EV) hosts the $\delta$~Sct stars in systems which are either eclipsing and single-line spectroscopic binaries or ellipsoidal and single or double-line spectroscopic binaries. For these systems, an assumption for one absolute parameter (e.g. the mass of the primary star) has been made with the others following it.

The unclassified systems, regardless of their $P_{\rm orb}$ value, are cases with unknown Roche geometry. Preventing reasons for their classification are: a) their light curves have not been obtained yet, b) their light curves are incomplete, c) their light curves are available but no analysis has been made, and d) the system is not eclipsing.

Systems with relatively long $P_{\rm orb}$ values (e.g. $>100$~days) can also be potentially considered as detached ones, but for reasons of homogeneity of the sample (i.e. lack of evidence) they are simply listed as unclassified. On the other hand, systems in eccentric and relatively wide (i.e. $P_{\rm orb}$ ranging from tenths to millions of days) orbits (marked with the superscript `$e$' in Tables~\ref{tab:tab1}-\ref{tab:tab2}) are considered directly as detached ones (e.g. BS~Aqr, IK~Peg). The latter assumption is based on the works of \citet{PET99} and \citet{DAV13}, who concluded that eccentric semidetached systems are rather rare and their majority have relatively short orbital periods (i.e. they range between 0.5-12 days).

\begin{figure}
\includegraphics[width=\columnwidth]{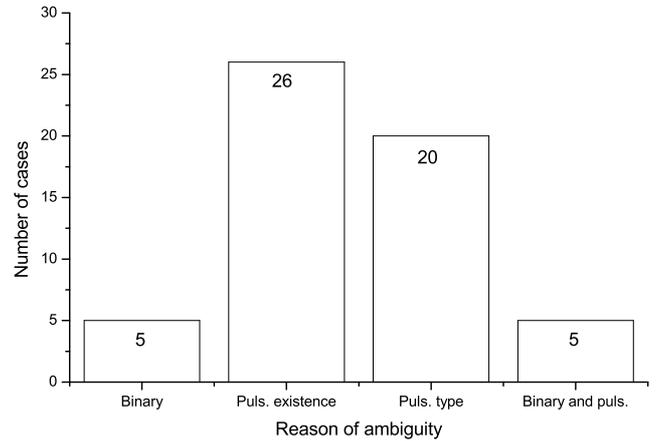}
\caption{The distribution of doubtful cases of binaries with a $\delta$~Sct component according to the reason of ambiguity. For details see Table~\ref{tab:tab4}.}
    \label{fig:stat3}
\end{figure}

According to our literature search, 56 cases are doubtful for being binaries with a $\delta$~Sct member. In particular, for 5 cases the binarity has not yet been proven, for 26 systems the pulsations occurrence is doubtful, 20 cases are indeed binaries that host a pulsator but the type of pulsations needs further investigation, and 5 cases are ambiguous concerning both the binarity and the existence of pulsations. The demographics of all these cases are shown in Fig.~\ref{fig:stat3}, while details for each individual are listed in Table~\ref{tab:tab4}.

On the other hand, another 42 cases are rejected from being members of this group (Table~\ref{tab:tab5}), either because they were proven to be single stars or they are binaries but: a) none of their components exhibit pulsations, b) their pulsations type is different than that of $\delta$~Sct stars, and c) they have properties other than pulsational (e.g. flare stars).

\begin{figure}
\includegraphics[width=\columnwidth]{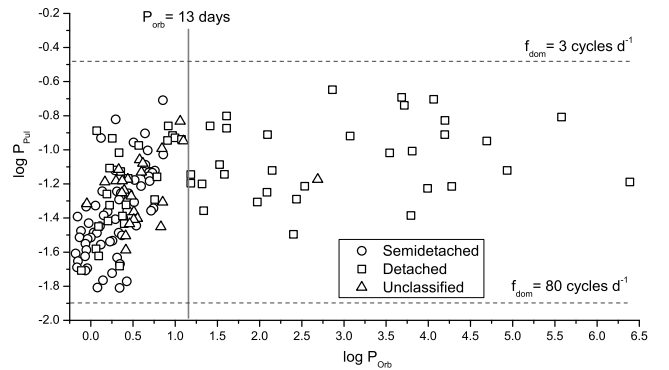}
\caption{The $P_{\rm orb}-P_{\rm pul}$ diagram of binary $\delta$~Sct stars. The limit of $P_{\rm orb}=13$~days is set for the separation of the two subgroups. Grey dash lines show the limits of the frequency range of $\delta$~Sct stars (3-80~cycle~d$^{-1}$).}
    \label{fig:statPP}
\end{figure}

Fig.~\ref{fig:statPP} shows the $P_{\rm orb}-P_{\rm puls}$ diagram for all of the currently known $\delta$~Sct stars in binaries. In this plot 147 out of 203 cases are included with their respective values to be given in Tables~\ref{tab:tab1} and \ref{tab:tab2}. 56 binary $\delta$~Sct stars are omitted because either their $P_{\rm pul}$ or the $P_{\rm orb}$ of their systems have not been determined so far (see Section~\ref{sec:s2} for details). The most important note that can be made for this diagram is the obvious existence of two subgroups. The first one includes the cases with the smaller values of $P_{\rm orb}$, while the second one those with larger $P_{\rm orb}$. The limit that has been set at $P_{\rm orb} \sim 13$~days is arbitrary, it may be apparent and its exact value needs further clarification. However, the strong linear correlation between $P_{\rm orb}-P_{\rm pul}$ for the systems with $P_{\rm orb}<13$~days is obvious even by naked eye, while for those with larger $P_{\rm orb}$ values these quantities seem to be uncorrelated. More systems with $10<P_{\rm orb}<30$~days are needed for a better estimation of this boundary. With more data points in this area, maybe this limit can reach the value of $P_{\rm orb}$=30-40~days (i.e. $\log P_{\rm orb}\sim1.5-1.6$). Therefore, for the correlations presented in the next section, only the properties of systems with $P_{\rm orb}<13$~days are taken into account.

\begin{landscape}
\begin{table}
\caption{List of binaries with $P_{\rm orb}<13$~days containing a $\delta$~Sct component.}
\label{tab:tab1}
\scalebox{0.79}{
\begin{tabular}{lccc cccc cccc cl}
\hline
Name	&	$P_{\rm orb}$	 &	$f_{\rm dom} $	&	$A$	&	ToV&	$M_{\rm pul}$	&	$R_{\rm pul}$	&	$T_{\rm pul}$	&	$\log g_{\rm pul}$	&	$M_{\rm comp}$	&	$R_{\rm comp}$	&	$a$	&	$q$	&	References\\
	    &	     (days)	     &	(cycle~d$^{-1}$)	&(mmag)	&		 &	(M$_{\sun}$)	&	(R$_{\sun}$)	&	(K)	            &(cm~s$^{-2}$)	&	(M$_{\sun}$)	&	(R$_{\sun}$)	&(R$_{\sun}$)	&	    	&	  \\
\hline
\multicolumn{14}{c}{Semi-detached systems}\\
\hline
J050634.16-353648.4$^{cat1}$ &	5.104238(5)&	13.45	&	65(7) ($V$)  &	SB2+E	&	1.73(11)&	2.40(7)&	7000*(200)	&	3.91(2)&	0.41(3)&	4.2(1) &	16.1(3)	&	0.236(4)&	NOR16                       \\
CZ Aqr	                     &	0.862752(3)&35.508(2)	&	3.7(5) ($B$) &	E   	&	2.0*     &	1.9(1)	&	8200*     	&	4.2(1)	&	0.98(1)&	1.8(1)	 &	5.62(1)	&	0.49(1) &	LIA12	                    \\
DY Aqr                       &	2.159679(1)&19.5645(4)  &14.1(7) ($V$)	 &	SB1+E	&	1.8(2)	 &	2.1(1) &	7625(125)  &	4.1(1) &	0.55(4)&	2.7(1)	 &	9.4(5)     &	0.31(2) &	SOY09, ALF14	            \\
RY Aqr      	             &	1.966565(1)&  6.62(8)   &32.7(4) ($B$)	 &	SB2+E   &	1.27(7) &	1.39(7)	&	7650*	    &	4.3(6)	&	0.26(2)	&	1.9(1)	 &	7.6(1)	&	0.204(7) &	POP89, MAN16	            \\
QY Aql	                     &	7.229560(1)&10.6561(1)  &11.8(2) ($B$)	 &	E	    &	1.6*(2) &	4.1(2)	&	7300*	    &	3.4(1)	&	0.4(1)	&	5.4(2)	 &	20.3(5)	&	0.25(2) &	LIA12, LIA13b	            \\
V729 Aql                     &	1.281905(1)&28.034(1)	&	4.2(4) ($W$) &	E	    &	1.5*(2) &	1.96(1)&	6900*    	&	4.0(1)	&	0.7(1)	&	2.03(1) &	6.6(1) 	&	0.44(1) &	LIA14	                    \\
V1464 Aql                    &	0.69777	    &24.621(1)	& 17.7(3) ($B$)	 &	SB1+EV	&	2.2(4)	 &	2.5(6)	&	7100*	    &	4.0(2)	&	0.3(1)	&	1.0(2)	 &	4.6(1)	    &	0.15(1) &	POU04, DAL13, LIA13a        \\
XZ Aql 	                     &2.1392073(7)	&30.631(1)	& 5.4(4) ($B$)	 &	SB2+E	&	2.42(14)&	2.45(7)&	8770	    &	4.04(6)&	0.45(6)&	2.43(6) &	9.9(2)	    &	0.184(4)&	SOY16, ZOL16	            \\
EW Boo 	                     &0.9063521(3) &49.4496(1)  &21.3(5) ($V$)	 &	SB2+E	& 2.00(4)   &1.88(2)  &	8970*	    &4.19(1)	&0.327(8)	&	1.23(1) &	5.22(4)    &	0.164(1)&	SOY08, SOY10, ZHA15a, DOG15 \\
YY Boo 	                     &3.93305(1)   &16.31828(2) & 58.4(2) ($B$)  &	E	    &		     &		    &	8000*       &		    &		    &		     &		        &	0.29(1) &	HAM10	                    \\
Y Cam 	                     &3.305766(3)  &15.0469(1)  & 8.2(2) ($V$)   &	SB2+E	&	2.08(9) &	3.14(5)&	8000(250)	&	3.76(2)&	0.48(3)&	3.33(5) &	13.1(2)    &	0.23(2) &	KIM03, SOY06b, ROD10, HON15 \\
R CMa 	                     &1.1359536(7)	&21.231	    &	8.8 ($B$)	 &	SB2+E	&	1.67(8) &	1.78(3)&	7300*	    &	4.16(2)&	0.22(7)&	1.22(7) &	5.8(3)	    &	0.13(4) &	MKR00, SOY06b, BUD11	    \\
TY Cap	                     &	1.423458(2)&24.222(1)	&18.5(7) ($B$)	 &	   E	&	2.0*	 &	2.5(1)	&	8200*	    &	3.9(1)	&	1.05(1)&	2.5(1)	 &	7.09(3)	&	0.52(1) &	LIA09, LIA12	            \\
AB Cas	                     &1.3668918(2)	&	17.492	&	26.6 ($B$)	 &	   E    &  1.87(17) &1.90(6)  &	8000*	    &	4.16(4)&	0.38(3)&	1.63(5) &	7.0(1)	    &	0.20(2) &	RO04, SOY06b, ABE07, KHA12  \\
IV Cas	                     &0.9985078(2)	&32.692(1)	&	3.4(2) ($B$)&	    E	&	1.98(10)&	2.13(4)&	8500	    &	4.08(3)&	0.81(4)&	1.80(3) &	6.06(2)	&	0.408(1)&	KIM05d, KIM10	            \\
RZ Cas	                     &1.195241(8) &64.1965(7)&	2.7(3) ($B$)&	SB2+E	&	2.01(2) &	1.61(1)&	 8907(15)	&4.328(4)	&	0.69(1)&	1.93(3) &	 6.59(3)	&	0.342(2)&  OHS01, SOY06a, SOY07, GOL07, TKA09 \\	
V1264 Cen$^{a1}$ 	         &5.3504751(6)	&13.62(2)  &	350.0 ($V$)	 &	SB2+E	&	1.49(2) &	2.35(2)&	7500	    &	3.87(1)&	0.33(2)&	4.04(1) &	16.1(3)	&	0.22(2) &	CHR07	                    \\
XX Cep                       & 2.33736(7)  &	31.77	&	2.9(2) ($B$)&	SB2+E	&	2.49(6) &	2.27(2)&	7946	    &4.12(1)	&	0.38(1)&	2.43(2) &	10.5(1)	&	0.151(2)&	ANG06, LEE07, HOS14, KOO16 	\\
WY Cet	                     &	1.939689(1)&13.211(1)	&7.7(3) ($B$)	 &	   E	&	1.7*	 &	2.2(1)	&	7500(150)	&	4.0(1)	&	0.44(1)&	2.3(1)	 &	8.7(3)	    &	0.26(1) &	LIA09, LIA12	            \\
UW Cyg	                     &	3.450759(1)&27.841(2)	&1.9(2) ($B$)	 &	   E	&	1.9*	 &	2.2(1)	&	8000(150)	&	4.0(1)	&	0.26(1)&	2.9(1)	 &	12.7(2)    &	0.14(1) &	LIA12	                    \\
V346 Cyg                     &	2.7433338	&	19.912	&	30.0 ($B$)	 &	   E	&	2.3	     &	3.8	    &	8353        &	3.6	    &	1.83	&	 4.74	 &	 13.55	    &	  0.8	 &	BUD04, KIM05b, SOY06b, HAN14\\
V469 Cyg                     &1.3124995(8) &	35.971	&	20.0 ($V$)	 &	   E	&		     &		    &		        &		    &		    &	         &		        &		     &	CAT04	                    \\
BW Del                        & 2.423133(6) &  25.100(1) &   29(2) ($B$) &    E      &    1.5*(2) & 2.1(1)   &  7000*        &   4.0(1)  &  0.3(1)   &   2.2(1)   &     9.3(4)   &   0.16(2)  &  LIA13b                       \\
GQ Dra 	                     &0.7659023(4) &29.860(1)	&	2.2(5) ($B$)&	   E	&		     &		    &		        &		    &		    &		     &		        &			 &private commun.       \\
HL Dra 	                     &0.9442760(2) &26.914(1) &3.0(2) ($B$)	 &	   E    &	 2.5(2) &	2.5(4)	&	  8200	    &	4.0(1)	&	0.9(1)	&	1.8(3)	 &	 6.1(2)	&	0.37(1) &	  LIA12	                    \\
SX Dra 	                     &5.169412(1)  &22.7423(1)& 17.3(4) ($B$) &	   E	&	 1.75*	 &	2.42(7)&	  7762*	    &	3.91(4)&	0.75(5)&	 5.36(5)&	   17.5(1) &	0.430(1)&	DIM10, SOY13	            \\
TW Dra 	                     &2.806791(3)	&	17.986	&	10.0 ($B$)	 &	SB2+E	&	2.2(1)  &	2.64(4)&	7815(92)	&3.93(3)	&	0.93(5)& 3.66(6)   &	12.48(2)	&	0.430(2)&	KIM03, ZEJ10, TKA10, BOZ13  \\
TZ Dra 	                     &	0.866033(1)&50.993(2)	&3.7(2) ($B$)	 &	   E	&	1.8*(2) &	1.7(1)	&	  7800*	    &	4.2(1)	&	0.6(1)	&	1.5(1)	 &	  5.2(2)	&	0.31(3) &	SOY06b, LIA13b	            \\
WX Dra$^{a2}$                & 1.801863(3) &34.0782(1)&1.171$^k$(1) ($Kp$)&   E	&		     &          &   7800*	    &	3.918*  &   	    &		     &              &	         &	PET12, this paper           \\
AS Eri	                     &2.664145(3) 	&59.03116(5)&	3.4(4) ($V$)&	   E	&	 1.93*	 &	1.57*	&	8470*       &	 4.3    &		    &		     &		        &		     &	MKR04, NAR13	            \\
TZ Eri	                     &	2.606113(3)&18.7174(3)&8.3(1) ($B$)	 & SB2+E	&	1.97(6) &	1.69(3)&  9307(20)	&	4.28(3)&	0.37 (1)&	2.60(4) &	10.6(2)	&	0.177(1)&	BAR98, LIA08	            \\
GSC 3889-0202	             &	2.71066(8)	&	22.676	&	50.0 ($V$)	 & SB1+E	&	         &		    &	  7750	    &	3.9	    &		    &		     &		        &		     &	DIM08b	                    \\
GSC 4293-0432$^{a3}$	     &	4.3844(2)  &	8.0	    &	40.0 ($B$)	 &	SB1+E	&	         &		    &	  7750	    &		    &		    &		     &		        &		     &	DIM09b	                    \\
GSC 4588-0883	             &	3.25855(9)	&	20.284	&	15.0 ($R$)	 &	SB1+E	&	         &		    &	7650	    &	3.9	    &		    &		     &		        &	0.16	 &	DIM09a	                    \\
BO Her 	                     &	4.272834(2)&13.430(1)	&68.0(3) ($B$)	 &	  E	    &	 1.8* (2)&	2.5(1)	&	  7800*	    &	3.9(1)	&	0.4(1)	&	3.8(1)	 &	14.8(1)	&	0.22(2) &	SUM07, LIA13b	            \\
CT Her 	                     &1.7863793(5)	&52.9366(1)&3.3(1) ($B$)  &  SB1+E   &  2.28(1)   &	2.06(6)&	  8200*	    &	4.17(2)&	0.29(4)&	1.87(8) &	8.69(3)	&0.1267(6)  &	KIM04b, LAM11	            \\
EF Her 	                     &4.729170(3)	&	10.07	&	69.0 ($B$)	 &	    E	&	 1.82*	 &	 2.8	&	 7300*	    &	3.8	    &	 0.27	&		     &	 15.14	    &	0.15	 &	KIM04b, SEN08, SOY13	    \\
LT Her           	         &1.0840335(4)	&30.521(3)	&	6.0 ($B$)	 &	   E	&      		 &		    &	9400*	    &		    &		    &		     &		        &		     &	LIA13b, STR16     \\
TU Her 	                     &2.266944(2)	&	18.0(2)&	5.0 ($V$)	 &	   E	&		     &		    &		        &		    &		    &		     &		        &		     &	LAM04	                    \\
RX Hya 	                     &2.2816596(9) 	&	19.38	&	7.0 ($B$)	 &	SB1+E	&	 2.72	 &	2.17	&	8770	    &	4.2	    &	0.66	&   2.93	 &	  11.2	    &	0.243	 &	LUC71, VYA89, KIM02, KIM03	\\
KIC 06220497                 & 1.32317(3)  &8.5175(3) &	4.6(2) ($Kp$) &  E     &	1.60*(8)&	2.69(6)&	7270*(6)	&	3.78(3)&	0.39(2)&	1.68(4) &	6.53(2)	&	0.242(3)&	GAU14, LEE16a	            \\
KIC 06669809                 &0.7337378(7) &32.5636(2)&0.560$^k$(1) ($Kp$)&	E	&	1.8*(3)	     &	2.5*(9) &	7462*(261)	&	3.9*(4)  &		    &		     &		        &		     &	HUB14, LIA16	       \\
KIC 10619109                 &2.045183(3)  &42.801(1)	&0.606$^k$(4) ($Kp$)&	E	&	1.5*(3)	 &	2.1*(8) &	7138*(284)	&	3.9*(4)  &		    &		     &		        &		     &	HUB14, this paper	      \\
KIC 11175495                 & 2.191027(3) &64.4434(1)&3.016$^k$(5) ($Kp$)&	E	&	2.0*(3)	     &	3.1*(5)	    &	8293*(290)	    &	3.8*(4)  &		    &		    &		      &	     &	HUB14, this paper	       \\
AU Lac                       &1.3924354(3)	&58.217(1) &  5.0(3) ($B$)  &	  E	    &	2.0*     & 1.8(1)  &	  8200*	    &	4.2(1)	&	0.6(1)	&	 2.1(1) &	   7.4(2)  &	 0.30(1)&	LIA12	                    \\
WY Leo 	                     &4.9858404	    &15.2528(1)&  11(1) ($V$)	 &	  E	    &	2.31	 &    3.26	&	  9640	    &	 3.78	&	  1.41	&	  2.65   &        19.02	&	    0.61 &	BRA80, DVO09	            \\
Y Leo 	                     &1.686086(1)	&34.4834(6)&4.1(2) ($V$) &	  E	    &	2.29	 &	1.90	&	  8700	    &	4.24	&	0.74	&		     &	   8.63	    &	   0.32	 &	TUR08, TUR11, POP11	        \\
RR Lep	                     &0.9154229(3) &33.2802(4)&	9.6(4) ($B$)&	  E	    &	1.8*(2) &	2.2(1)	&	  7800*	    &	4.0(1)	&	0.4(1)	&	1.4(2)	 &	  5.3(2)	&	0.23(2) &	DVO09, LIA13b	            \\
CL Lyn	                     &	1.58606	    &23.051(1)	&	7.3 (3) ($B$)&    E	    &	2.0*     &	2.5(1)	&	  8200*	    &	3.9(1)	&	0.38(3)&	1.9(1)	 &	  7.9(2)	&	0.19(2) &	LIA12	                    \\
VY Mic	                     &4.436373(2)	&12.23411(3)&19.4(2) ($V$) & E	    &	2.39	 &	2.24	&	   8500     &	4.12	&	1.96	&	 4.43	 &	  19.01   	&	0.82	 &	BUD04, PIG07	            \\
V2365 Oph                    &	4.86560(1)	&	14.286	   &	50.0 ($V$)	&SB2+E	&1.97(2)	 &2.19(1)	&	9500(200)	&4.051(3)	&1.06(1)	&	0.934(4)&	17.46(5)	&	0.538(3)&	IBA08	                    \\
FR Ori                       &0.8831653(2)	&	38.638	   &	5.8 ($V$)	&	E	&		    &		    &	    7830*	&	    	&		    &	         &        		&	0.325(2)&	YAN14	                    \\
V392 Ori                     &0.6592821(4)	& 40.575(7)   &2.4(2) ($B$)	&SB1+E	&	2.0*(2)&	2.04(7)&	   8300*	&	4.12(5)&	0.49(5)&	1.15(4) &	 4.3(1)	&	0.252(2)&	NAR02a, ZHA15b	            \\
MX Pav	                     &	5.730835(4)&13.22722(1)  &76.9(3) ($V$)  &	E	&		    &		    &	  8200	    &		    &		    &		     &		        &	         &	PIG07	                    \\
BG Peg	                     &	1.952404(1)& 25.544(1)   &13.1(6) ($B$)  &	E	&	2.5*	&	2.85(3)&    9000*	    &	3.93(1)&	0.55(3)&	2.37(3) &	9.76(8)    &	0.219(1)&  SOY09, DVO09, SOY11, LIA11, SEN14 \\
\hline
\end{tabular}}
\end{table}
\end{landscape}

\begin{landscape}
\begin{table}
\contcaption{   }
\scalebox{0.8}{
\begin{tabular}{lccc cccc cccc cl}
\hline
Name	&	$P_{\rm orb}$	 &	$f_{\rm dom} $	&	$A$	&	ToV&	$M_{\rm pul}$	&	$R_{\rm pul}$	&	$T_{\rm pul}$	&	$\log g_{\rm pul}$	&	$M_{\rm comp}$	&	$R_{\rm comp}$	&	$a$	&	$q$	&	References      \\
	    &	     (days)	     &	(cycle~d$^{-1}$)	&(mmag)	&		 &	(M$_{\sun}$)	&	(R$_{\sun}$)	&	(K)	            &(cm~s$^{-2}$)	&	(M$_{\sun}$)	&	(R$_{\sun}$)	&(R$_{\sun}$)	&	    		&       \\
\hline
AB Per	                     &7.16023(2)	&	5.116	   &	20.0 ($B$)	&SB2+E	&	1.9	    &	2.0  	&	8200*	    &	4.1   	&	0.23	&	 4.2	 &	20.58	    &	  0.12	 &  KIM03, KIM04a, BUD04,       \\
                             &              &              &                &          &            &          &               &           &           &            &              &         &  KIM06, SMA14                \\
IU Per	                     &0.8570201(1)	&	42.103	   &	20.0 ($B$)	&	E	&	2.2*	&	2.04	&	  8450*	    &	4.16	&	0.6	    &	1.46	 &	   5.35	    &	0.273(5)&	KIM05c, SOY06b, ZHA09, KUN13 \\
AO Ser	                     &0.8793463(3)	&	21.505	   &	20.0 ($B$)	&	E	&	1.82*	&	1.56	&	7800*	    &	4.3	    &	0.428	&	1.26	 &	5.055	    &	0.235(3)&	KIM04c, YAN10, ALT12        \\
UZ Sge	                     &2.215753(2)	&	46.652(6) &3.1(3) ($B$)	&	E	&	2.1*(2)&	1.9(2)	&	8700*   	&	4.2(1)	&	0.29(2)&	2.2(2)	 &	9.8(8)	    &	0.14(1) &	LIAN12	                    \\
AC Tau	                     &2.043420(7)	&	17.533(1) &6.0(1.0) ($V$)	&	E	&	1.45	&	2.30	&	7162	    &	3.88    &	0.99	&	 2.90    &	   9.33	    &	0.68     &	BUD04, SOY06c, DVO09	    \\
IZ Tel	                     &4.880219(4)	&13.55801(2)   &45.9(4) ($V$)  &	E	&		    &		    &	  	        &		    &		    &		     &       		&	         &	PIG07	                    \\
IO UMa	                     &5.5201878(1) &22.0148(1)   &6.5(2) ($B$)   &SB2+E	&	2.11(7)&	3.00(4)&	7800	    &	3.81(2)&	0.29(2)&	3.92(5) &	17.5(1)	&	0.135(3)&	LIA12, SOYD13	            \\
VV UMa                       &0.6873778(4)	&48.8483(3)   &2.8(1) ($V$)	&SB1+E	&	2.5*	&	1.77	&	9500*	    &	4.34	&	0.84	&	1.36	 &	    5.0	    &	0.337(2)&  STR50, LAZ01, KIM05a, GUN15, TAN15  \\
1200-03937339$^{cat2}$	     &	1.17962(1)	& 30.668(1)   &5.1(4) ($W$)	&	E	&	1.6(2)*	&	2.24(4)&	7250*	    &	3.9(1)	&	0.3(1)	&	1.44(3) &	 6.0(1)	&	 0.19(2)&	LIA14	                    \\
BF Vel                       &0.7040277(1)	&	44.94(2)  &2.6(2) ($B$)	&	E	&	1.98*	&1.77(1)	&	   8550*	&	4.23(4)&	0.84(8)&	1.481(6)&	4.816(6)	&	0.424(2)&	MAN09	                    \\
\hline
\multicolumn{14}{c}{Detached systems}\\
\hline
V551 Aur                     &1.173203(3)	&7.7270(7)	   &19.2(3) ($V$)  &	E	&		    &		    &	7000*	    &		    &		    &		     &		        &	0.725(6)&	LIU12	                    \\
V389 Cas                     &2.4948239(5)	&	27.1(8)    &	8.8 ($R$)   &	E	&		    &		    &	7673*(31)	&		    &		    &		     &		        &		     &	KOR15	                    \\
RS Cha$^{b1, e}$ A           &1.6698822(8)	&  21.11       &	            &SB2+E	&	1.89(1)&	2.15(6)&7638(176)	& 4.05(6)   &	1.87(1)&	2.36(6) &	9.20(3)	&	0.970(7)&	JOR75, MCI77, ALE05,        \\
RS Cha$^{b1, e}$ B           &              &   12.81      &                &       &   1.87(1)&	2.36(6)& 7228(166)  & 3.96(6)   &           &            &              &            &  BOH09, WOO13                \\
CoRoT 105906206	             &3.6945708(1) &9.4175(1)    &2.55(1) ($CoR$)  &SB2+E	&	2.25(4)&	4.24(2)&	6750(150)	&	3.53(1) &	1.29(3)&	1.34(1) &	15.32(8)	&	0.574(8)&	SIL14	                    \\
AK Crt$^{b2}$	             & 2.778758(4)	&14.71370(2)  &8-35 ($V$)	    &	E	&		    &		    &	7800	    &		    &		    &            &      		&		     &	PIG07	                    \\
GK Dra$^{e}$ 	             & 9.9742(2)	&8.4907(1)	  &	40 ($V$)	&SB2+E	&	1.8(1)	&	2.83(5) &	6878(57)   &	3.79(4)	&  1.46(7)  &  2.43(4) &28.9(4) &0.81(6)	&	DAL02, ZWI03, GRI03	\\
HN Dra$^{b3, e}$ 	         &1.80075(3)	& 8.5583(3)   &8.9(7) ($B$)	&SB2+EV	&	1.87*	&	2.87	&	6920*	    &	3.84	&	1.30	&	1.42	 &	9.36        &	0.690(7)&	CHA04                       \\
HZ Dra	                     &0.77293968	& 51.068(2)   &4.0(4) ($B$)	&SB1+E	&	3.0(3)	&	2.3(1)	&	9800*	    &	4.2(1)	&	0.4(1)	&	0.8(1)	 &	5.3(1)	    &	0.12(4) &	LIA12	                    \\
OO Dra$^{b4}$ 	             &1.2383777   	&41.867(1)    &4.9(3) ($B$)	&SB1+E	&1.97(25)	&	2.04(9)&	  8500*	    & 4.11(7)	&	0.19(3)&	1.17(5) &	6.42(6)	&	0.097(2)&	DIM08a, ZHA14	            \\
HD 172189$^{b5, e}$          & 5.70165	    &	19.5818(3)&	 	        &SB2+E	&2.06(15)	& 4.01(9)	&	7920(190)	&3.55(1)	&1.87(14)	& 2.97(7)   &	21.2(5)	& 0.91 (4)   &	MAR05, COS07, CRE09, IBA09 	    \\
HD 220687$^{b6}$	         &1.594251(3)	&26.16925(4)  &12.8(1) ($V$)  &	E	&		    &		    &	9000*	    &		    &      		&            &        		&		     &	PIG07	                    \\
V644 Her$^{b7, e}$           &11.85859(4)  &	8.688	   &	40.0 ($V$)	&	SB1	&	    	&		    &        		&    		&	     	&      	     &       		&	         &	BAR82, ROD00, DUC11	        \\
HIP 7666$^{b8}$	             &2.37229(8)	&	24.465(8) &	1.7($V$)	&	E	&		    &		    &	8200*	    &		    &		    &		     &		        &		     &	ESC05	                    \\
AI Hya$^{e}$                 &8.289649(7)	&	7.24478	   &	7.6 ($B$)	&SB2+E	&	1.98(4)&	2.77(2)&	7100(60)	& 3.850(4) &	2.15(4)&	3.92(3) &	28.30(9)	&1.08(1)	 &	JOR78, POP88, EKE14         \\
KIC 04544587$^{e}$           &2.189097(3)  &48.02231(4)  &0.329$^k$(4) ($Kp$)&SB2+E& 1.98(7)& 1.82(3)  &  8600(100)	& 4.241(9) &	1.61(6)&	1.58(3) &   10.86(5)  &  0.813(2) &	HAM13	                    \\
KIC 06629588	             & 2.264471(3)  &13.39649(1)  &3.05$^k$(1) ($Kp$) &E  &   1.2*(3)        &	1.8*(7)        &	6787*(247) 	    &	4.0*(4)	&		    &		     &       		&	         &	HUB14, LIA16 \\
KIC 09851944$^{b1}$ A&2.16390189(8) &10.399692(2)  &0.653$^k$(8) ($Kp$) &SB2+E &   1.76(7)   &	 2.27(3)     &	7026(100) 	  &	3.96(3)	&	1.79(7)	&	3.19(4)	 &    10.7(1)  &	 1.01(3)  &	GUO16 \\
KIC 09851944$^{b1}$ B&              &             &                 &      &    1.79(7)  &	 3.19(4)     &	 6902(100)	  &	3.69(3)	&		    &		     &      	   &	          &	      \\
KIC 10661783	             &1.231363(1)  & 28.135	   &4.163(8) ($Kp$) &	SB2+E&2.10(3)  &  2.58(2)	&   7764(49)   & 3.938(4) &0.191(3)  &  1.12(2)  &	6.52(6)	&	0.091(2)&	SOU11, LEH13	            \\
KIC 10686876	             &2.618427(4)  & 21.0243(2)  &0.289$^k$(2) ($Kp$) &E	&	 1.9*(2)      &   2.4*(8)		&	8167*(285)	    &	3.9*(4)	&	     	&	         &       &	     &	GIE12, HUB14, this paper \\
KIC 11401845	             &  2.1613(2)  &	13-24.5	   &		      &	E	&	1.7*(2)	    &	2.3*(7)	    &	7813*(273)       &	3.9*(4)  &		    &		     &		        &            &	HUB14, GAU14	     \\
V577 Oph$^{a4, e}$           &6.079089(5) 	&	14.3906(1)&	28.9 ($V$)	&SB2+E	& 1.60*(1)	&	        &		        &		    &	1.50(2)&		     &	            &0.939(6)	 &  VOL90, DIE93, ZHO01a, VOL10, CRE10 \\
FL Ori	                     &1.550972(1)	&	18.178	   &	44.0 ($V$)	&	E	&	2.90	&	2.10	&	    8700	&	    4.3	&	1.93	&	2.17	 &	  9.76	    &	0.67	 &	BUD04, ZAS11	            \\
XX Pyx	                     &	1.151145	&	38.110	   &10.1(3) ($B$)	&SB1+EV	&		    &		    &	8500*	    &		    &		    &		     &		        &		     &	ARE01, HAN02, AER02	        \\
V5548 Sgr$^{b9, e}$	     &	8.115800    &	8.818	   &	17.1 ($B$)	&	SB1	&	1.69*	&		    &     7600*		&		    &		    &		     &              &		     &	WAL69, HIL92, DUC11	        \\
18 Vul$^{b10, e}$            &9.31408(4)	&	8.2305	   &	10.0 ($V$)	&	SB2	&	2.4*	&	3.5(3)	&	8300*(300)	&	3.73(7)&	2.2*	&	2.4(2)	 &	 31.7(1)   & 0.91(2)   &	ROD00, FEK13	            \\
\hline
\multicolumn{14}{c}{Unclassified systems}\\
\hline
KW Aur$^{c1}$ 	             &3.7886(2)	&11.431 (8)    &	80.0 ($V$)	&	SB1	&	2.36(6)&	     	&	7800	    &	3.4*    &		    &		     &		        &		     &	HUD71, MOR76, ZOR12, GAL12      \\
RS Gru	                     &	11.49425	&	6.80218	   &	600.0 ($V$)	&	SB1	&		    &		    &	 7500*	    &		    &		    &		     &	            &		     &	NAM76, DER09, GAR12	        \\
HD 061199$^{c2}$	         &	3.5744(3)  & 25.256(1)   &1.46(5) ($V$)	&	SB2	&	2.1*    &	2.4*	    &	8000(200)	&	4.0(5)*&	2.03    &		     &		        &	0.968(4)&	HAR08	                    \\
HD 207651$^{c3}$             &1.470739(6)  &	15.43(1)  &21(2) ($B$)    &SB2+EV	&	1.9*	&		    &	  8200	    &		    &	      	&		     &		        &        	 &	HEN04, FEK15	            \\
TT Hor	                     &2.608215(3)	&	38.7(4)   &	20(4) ($V$)&	E	&		    &		    &		        &		    &    		&            &        		&		     &	MOR13	                    \\
KIC 04570326	             &1.121553(1)  &	8.5-20	   &		        &	EV	&	1.5*(2)	    &	2.1*(6)	    &	7000*(140)    &	4.0*(2)	    &		    &		     &		        &		     &	HUB14, GAU14  \\
KIC 04739791	             &0.8989(2)    &20.7389(1)  &	1.97 (5) ($Kp$)&  E &1.75*(8)  &	2.46*(9)    &	     7761*(271)	&	3.9*(4)  &		    &		     &		        &            &	HUB14, LEE16b	           \\
KIC 05197256	             &6.9634(2)    &9.847(2)     &         &  	E	&	1.9*(2)   &	  2.8*(6)  	&    7832*(274)     	& 3.9*(4)    &	    	&      		 &		        &		     &	HUB14, TUR15	        \\
KIC 05783368	             &1.8639(5)    &	8.5-22     &	            &   E	&	1.8*(2) &	2.8*(5) &	8133*(284)  & 3.8*(4)    &		    &	         &		        &		     &	HUB14, GAU14                 \\
KIC 05872506	             &2.126109(6)  &8.5-24.5      &                &    EV	&	1.6*(3) &	2.5*(7)	  &	7794(244)*  & 3.9*(3)	&     		&            &        		&     	     &	HUB14, GAU14	             \\
KIC 06541245	             &	1.6	         &19-24.5	   &		        &	EV	&		    &		    &	6545*(229)	    &	4.2*(4)	&		    &		     &		        &		      & HUB14, GAU14	         \\
KIC 11973705$^{c4}$	         &  6.77214(2)  &	28.260	   &		        &SB2+EV	& 1.6(3)	&	 2.0*  	&	7404*	    &	4.039*	&		    &		     &		        &		      &	LEH11, BAL11, CAT11         \\
DG Leo$^{c5}$	             &4.146751(5) 	 &	11.994	   &     6.2 ($B$)  &SB2+EV	&	2.0(2)	& 2.96(20) &	7470(220)  &3.8(1)	&	2.0(2)	&	2.94(20)&	17.65(1)	&	1.000(1) &	LAM05, FRE05	            \\
CQ Lyn	                     &	12.50736(8) &	8.8673     &	40.0 ($V$)	&	SB2	&1.71(6)	&2.6(3)	&	6760(50)	&3.85(8)	&1.07(4)	&	1.0(1)&	32.7(1)   &0.626(4)  &	CAR02	                    \\
\hline
\end{tabular}}
\end{table}
\end{landscape}

\begin{landscape}
\begin{table}
\contcaption{   }
\scalebox{0.8}{
\begin{tabular}{lccc cccc cccc cl}
\hline
Name	&	$P_{\rm orb}$	 &	$f_{\rm dom} $	&	$A$	&	ToV&	$M_{\rm pul}$	&	$R_{\rm pul}$	&	$T_{\rm pul}$	&	$\log g_{\rm pul}$	&	$M_{\rm comp}$	&	$R_{\rm comp}$	&	$a$	&	$q$	&	References   \\
	    &	     (days)	     &	(cycle~d$^{-1}$)	&(mmag)	&		 &	(M$_{\sun}$)	&	(R$_{\sun}$)	&	(K)	            &(cm~s$^{-2}$)	&	(M$_{\sun}$)	&	(R$_{\sun}$)	&(R$_{\sun}$)	&	    		&	 \\
\hline
RZ Mic                       &  3.9830406(7) &13.52(4)   &19.0(6) ($V$)  &	E	&		    &		    &		        &		    &		    &		     &		        &		      &	STR16	                    \\
V937 Mon$^{c6}$	             &  3.20865(1)	 &9.05142(1)  &41.7(1) ($V$)  &	E	&		    &		    &		        &		    &		    &		     &		        &		      &	PIG07	                    \\
V1004 Ori$^{c7}$	         &	2.74050	     &	14.9(2)   &	12.3 ($B$)	&	SB1	&2.15*(2)	&2.78*(21)	&	7706*(30)	& 3.71*(6)	&		    &		     &		        &		      &	POU04, DEB09	            \\
GX Peg$^{c8}$	             &	2.3410	     &	17.857	   &	 	        &	SB1	&		    &		    &		        &		    &		    &		     &		        &		      &	MIC93, POU04, ABT04         \\
HM Pup	                     &  2.589685(3) &	32(1)     &  10(4) ($V$)	&	E	&		    &		    &	  7800*	    &		    &		    &		     &		        &		      &	MOR13	                    \\
LY Pup	                     &  2.887907   &	27.2(1)     &  20(1) ($V$)	&	E	&		    &		    &	     	    &		    &		    &		     &		        &		      &	STR16	                    \\
V632 Sco                     &  3.2204     &	25.6(1)     &  6(1) ($V$)	&	E	&	1.99    &	1.67    &	6170   	    &	4.29    &	0.96    &	0.25     &	8.28        &	0.48      &	BRA80, ALF12, STR16	        \\
V638 Sco                     &  2.3582287  &	15.2(2)     &  5(1) ($V$)	&	E	&	2.60    &	3.19    &	8900   	    &	3.85    &	1.25    &	0.58     &		11.69   &	0.48      &	BRA80, ALF12, STR16	        \\
V483 Tau$^{c9}$	             &	2.486(2)	 &	17.25689   &	2.07 ($V$)	&	SB1	&		    &		    &		        &		    &		    &		     &		        &		      & KAY99, PAPA00	            \\
V775 Tau$^{c10}$	         &	2.1436117(8)&	13.0364	   &	6.01 ($u$)	&	SB1	&		    &		    &		        &		    &		    &		     &		        &		      &	ZHI00a, GRI12	            \\
V353 Tel$^{c11}$ 	         &	3.24886	     &	23.0838	   &	15.3 ($V$)	&	EV	&		    &		    &		        &		    &		    &		     &		        &			  &	HANS02	                    \\
$\theta$ Tuc	             &	7.1036(5)   &	20.28061   &18.5(2) ($b$)  &	SB2	&2.1*(1)	&		    &	    7575*	&		    &		    &		     &		        &	0.090(2) &	MEY98, TEM00, PAP00	        \\
0975-17281677$^{cat2}$       &	3.0155(8)	 &	18.702(1) & 6.5(3) ($W$)	&	E	&		    &	     	&	        	&    		&	     	&            &           	&      		  &	LIA14	                    \\
AW Vel                       & 1.992475(2)  &	15.2(3)   & 58(1) ($V$)	&	E	&		    &		    &	  7800*	    &		    &   		&		     &		        &		      &	MOR13	                    \\
\hline
\end{tabular}}
\newline
*assumed, $^{cat1}$1SWASP catalogue, $^{cat2}$USNO A2.0 catalogue, $^{k}\times10^{-3}$ in relative flux units, $^{e}$eccentric orbit,
$^{a1}$UNSW-V-500, $^{a2}$KIC~10581918, $^{a3}$BD+65~1939, $^{a4}$triple system,
$^{b1}$both components (referred as A and B) are pulsators, $^{b2}$HD~099612, $^{b3}$HD~173977, $^{b4}$GSC~4550-1408, $^{b5}$BD+05~3864, $^{b6}$BD-12~6485, $^{b7}$HD~152830, $^{b8}$HD~232486, $^{b9}$h01~Sgr, HD~184552, $^{b10}$HD~191747,
$^{c1}$14~Aur, HR~1706, Multiple system $^{c2}$BD+05~1734, triple system, $^{c3}$HIP~107786, triple system, $^{c4}$SPB star+$\delta$~Sct or normal star+hybrid, $^{c5}$triple system, $^{c6}$HD~062571, $^{c7}$59~Ori, HD~40372, $^{c8}$HIP~111191, HR~8584, $^{c9}$57~Tau, $^{c10}$60~Tau, HD~27628, $^{c11}$HD 173794	\\
ABE07-\citet{ABE07}, ABT04-\citet{ABT04}, AER02-\citet{AER02}, ALE05-\citet{ALE05}, ALF12-\citet{ALF12}, ALF14-\citet{ALF14}, ALT12-\citet{ALT12}, ANG06-\citet{ANG06}, ARE01-\citet{ARE01}, BAL11-\citet{BAL11}, BAR98-\citet{BAR98}, BAR82-\citet{BAR82}, BOH09-\citet{BOH09}, BOZ13-\citet{BOZ13}, BRA80-\citet{BRA80}, BUD04-\citet{BUD04}, BUD11-\citet{BUD11}, CAR02-\citet{CAR02}, CAT11-\citet{CAT11}, CAT04-\citet{CAT04}, CHA04-\citet{CHA04}, CHR07-\citet{CHR07}, COS07-\citet{COS07}, CRE09-\citet{CRE09}, CRE10-\citet{CRE10}, DAL13-\citet{DAL13}, DAL02-\citet{DAL02}, DEB09-\citet{DEB09}, DER09-\citet{DER09}, DIE93-\citet{DIE93}, DIM08a-\citet{DIM5842}, DIM08b-\citet{DIM5856}, DIM09a-\citet{DIM5883}, DIM09b-\citet{DIM5892}, DIM10-\citet{DIM10}, DOG15-\citet{DOG15}, DUC11-\citet{DUC11}, DVO09-\citet{DVO09}, EKE14-\citet{EKE14}, ESC05-\citet{ESC05}, FEK13-\citet{FEK13}, FEK15-\citet{FEK15}, FRE05-\citet{FRE05}, GAL12-\citet{GAL12}, GAR12-\citet{GAR12}, GAU14-\citet{GAU14}, GIE12-\citet{GIE12}, GOL07-\citet{GOL07}, GRI03-\citet{GRI03}, GRI12-\citet{GRI12}, GUN15-\citet{GUN15}, GUO16-\citet{GUO16}, HAM13-\citet{HAM13}, HAM10-\citet{HAM10}, HAN02-\citet{HAN02}, HANS02-\citet{HANS02}, HAN14-\citet{HAN14}, HAR08-\citet{HAR08}, HEN04-\citet{HEN04}, HIL92-\citet{HIL92}, HON15-\citet{HON15}, HOS14-\citet{HOS14}, HUB14-\citet{HUB14}, HUD71-\citet{HUD71}, IBA08-\citet{IBA08}, IBA09-\citet{IBA09}, JOR75-\citet{JOR75}, JOR78-\citet{JOR78}, KAY99-\citet{KAY99}, KHA12-\citet{KHA12}, KIM02-\citet{KIM02}, KIM03-\citet{KIM03}, KIM04a-\citet{KIM04a}, KIM04b-\citet{KIM5537}, KIM04c-\citet{KIM04SER}, KIM05a-\citet{KIM05UMA}, KIM05b-\citet{KIM05346}, KIM05c-\citet{KIM05Per}, KIM05d-\citet{KIM05CAS}, KIM06-\citet{KIM06}, KIM10-\citet{KIM10}, KOO16-\citet{KOO16}, KOR15-\citet{KOR15}, KUN13-\citet{KUN13}, LAM04-\citet{LAM04}, LAM05-\citet{LAM05}, LAM11-\citet{LAM11}, LAZ01-\citet{LAZ01}, LEE07-\citet{LEE07}, LEE16a-\citet{LEE16a}, LEE16b-\citet{LEE16b}, LEH11-\citet{LEH11}, LEH13-\citet{LEH13}, LIA08-\citet{LIA08}, LIA09-\citet{LIA09}, LIA11-\citet{LIA11}, LIA12-\citet{LIA12}, LIAN12-\citet{LIAN12}, LIA13a-\citet{LIA13a}, LIA13b-\citet{LIA13b}, LIA14-\citet{LIA14}, LIA16-\citet{LIA16}, LIU12-\citet{LIU12}, LUC71-\citet{LUC71}, MAN09-\citet{MAN09}, MAN16-\citet{MAN16}, MAR05-\citet{MAR05}, MCI77-\citet{MCI77}, MEY98-\citet{MEY98}, MIC93-\citet{MIC93}, MKR00-\citet{MKR00}, MKR04-\citet{MKR04}, MOR76-\citet{MOR76}, MOR13-\citet{MOR13}, NAM76-\citet{NAM76}, NAR02a-\citet{NAR02a}, NAR13-\citet{NAR13}, NOR16-\citet{NOR16}, OHS01-\citet{OHS01}, PAP00-\citet{PAP00}, PAPA00-\citet{PAPA00}, PET12-\citet{PET12}, PIG07-\citet{PIG07}, POP11-\citet{POP11}, POP88-\citet{POP88}, POU04-\citet{POU04}, ROD00-\citet{ROD00}, ROD04-\citet{ROD04}, ROD10-\citet{ROD10}, SEN08-\citet{SEN08}, SEN14-\citet{SEN14}, SIL14-\citet{SIL14}, SMA14-\citet{SMA14}, SOU11-\citet{SOU11}, SOY06a-\citet{SOY06a}, SOY06b-\citet{SOY366}, SOY06c-\citet{SOY370}, SOY07-\citet{SOY07}, SOY08-\citet{SOY08}, SOY09-\citet{SOY09}, SOY10-\citet{SOY10}, SOY11-\citet{SOY11}, SOY13-\citet{SOY13DRA}, SOYD13-\citet{SOY13UMA}, SOY16-\citet{SOY16}, STR16-\citet{STR16}, SUM07-\citet{SUM07}, TAN15-\citet{TAN15}, TEM00-\citet{TEM00}, TKA09-\citet{TKA09}, TKA10-\citet{TKA10}, TUR08-\citet{TUR08}, TUR11-\citet{TUR11}, TUR15-\citet{TUR15}, VOL90-\citet{VOL90}, VOL10-\citet{VOL10}, VYA89-\citet{VYA89}, WAL69-\citet{WAL69}, WOO13-\citet{WOO13}, YAN10-\citet{YAN10}, YAN14-\citet{YAN14}, ZAS11-\citet{ZAS11}, ZEJ10-\citet{ZEJ10}, ZHA09-\citet{ZHA09}, ZHA14-\citet{ZHA14}, ZHA15a-\citet{ZHA15a}, ZHA15b-\citet{ZHA15b}, ZHI00a-\citet{ZHI00Tau}, ZHO01a-\citet{ZHO01a}, ZOL16-\citet{ZOL16}, ZOR12-\citet{ZOR12}, ZWI03-\citet{ZWI03}
\end{table}
\end{landscape}


\begin{landscape}
\begin{table}
\caption{List of detached binaries with $P_{\rm orb}>13$~days containing a $\delta$~Sct component.}
\label{tab:tab2}
\scalebox{0.82}{
\begin{tabular}{lccc cccc cccc cl}
\hline
Name	&	$P_{\rm orb}$	&	$f_{\rm dom} $	&	$A$	&	ToV&	$M_{\rm pul}$	&	$R_{\rm pul}$	&	$T_{\rm pul}$	&	$\log g_{\rm pul}$	&	$M_{\rm comp}$	&	$R_{\rm comp}$	&	$a$	&	$q$	&	References\\
	    &	     (days)	    &	(cycle~d$^{-1}$)	&(mmag)	&		 &	(M$_{\sun}$)	&	(R$_{\sun}$)	&	(K)	            &(cm~s$^{-2}$)	&	(M$_{\sun}$)	&	(R$_{\sun}$)	&(R$_{\sun}$)	&	    	&	  \\
\hline
BS Aqr$^{e}$ 	         &	   11578	&	5.055(99)&	184 ($V$)	 &O$-$C	&		    &		    &		       &		    &		     &		     &		    &		    &YAN93, ROD98a, FU99, FU00 	        \\
CY Aqr$^{e}$ 	         &	18956(37)   &16.38314(1) & 710 ($V$)	 &O$-$C	&		    &		    &		       &		    &		     &		     &		    &		    &	FU00, FU03, DER09, STE12	    \\
$\kappa^2$ Boo	         &2437977(88022)&	15.434	 &	7.8 ($V$)	 &	Vis	&	2.12*	&	2.78*	&	7875*	   &  3.88		&		     &		     &		    & 		    &	FRA95, MAL12	                \\
AI CVn$^{d1, e}$         &	124.44(3)	&  8.595	 &10.6(1) ($y$)	 & 	SB1	&		    &		    &	6875(120)  &	3.3(4)	&		     &		     &		    &		    &	BRE08, SCH14	                \\
AD CMi$^{e}$ 	         &	15661(303)  & 8.131768(2)&	143(1) ($V$) &O$-$C	&		    &		    &		       &		    &		     &		     &		    &     		&	FU00, KHO07, HUR07, DER09       \\
$\epsilon$ Cep 	         &	6209	    &	24.2471	 &	7.9 ($B$)	 &	Vis	&	1.8(2)	&	    2.0	&	7350	   &	4.09	&		     &		     &	      	&		    &	LOP79, HAR84, BAA93,            \\
                &              &              &                &          &             &           &              &            &           &            &          &           &   MAW11                           \\
$\delta$ Del$^{d2, e}$A	 &	 40.58	    &	6.3291	 &	 70 ($V$)	 &	SB2	&		    &		    &7334*		   &		    &		     &		     &		    &		    &	REI76, BAA93, ROD00, 	        \\
$\delta$ Del$^{d2, e}$B	 &	     	    &	7.4627	 &	        	 &		&		    &		    &		       &		    &		     &		     &		    &		    &	PRU11	                        \\
OR Dra$^{d3, e}$        &	 49271	    & 8.86297(3) &11.0(7) ($B$)	 &	Vis	&		    &		    & 7000*		   &		    &		     &		     &		    &		    &	HEN01, MAL12	                \\
HD 051844$^{d4, e}$	     &	33.498(2)	& 12.21284(4)&2.26(2) ($W$)  &SB2+EV&  1.97(19) &	3.52(17)&	7300(200)  &	3.65(1)	&	2.01(20) &	3.7(2)	 &	69.4(5)	&	0.99(1)	&	HAR14	                        \\
DY Her$^{d5}$           &15720(2620)   & 6.72806(1) &	 490 ($V$)	 &O$-$C	&		    &	2.8*(2)	&		       &		    &		     &		     &		    &		    &	YAN93, MIL94, POC00, POJ02	    \\
KZ Hya	                 &	9788(73)	&16.80378(1) &	 192 ($R$)	 &O$-$C	&		    &		    & 7300*(530)   &   4.0*(2)	&		     &		     &		    &		    &	ROD00, FU08	                    \\
KIC 03858884$^{d6, e}$   &25.9519(1)  &7.2306(1)	 &10.2(2) ($Kp$) &SB2+E &	1.88(3) &	3.45(1) &	6800(70)   &	3.63(1) &   1.86(4)	 &	3.05(1)  &	57.2 (2)&	0.988(2)&	CAT10, LEE12, MAC14	            \\
KIC 04150611$^{d7}$	     &94.1982(7)	&20.24360(3) &1.610(7) ($Kp$)&	E	&		    &		    &	   6623*   &	4.1*(4) &		     &		     &		    &		    &	SHI12, GRE13, HUB14             \\
KIC 08264492$^{e}$	     &	252.39(56)  &31.292032(2)&0.893(3) ($Kp$)&O$-$C	&	1.87*(2)&	2.4(9)  &	 7991(279)*&	4.0(4)*	&		     &		     &		    &		    &	HUB14, SHI15, MUS15	            \\
KIC 08569819$^{d8, e}$	 &20.84993(3)	&15.857472(1)&4.150(5) ($Kp$)&	E	&	1.7*	&	     	&	7100(250)  &		    &	    1.0  &	         &	  44.6	&	0.588	&	KUR15	                        \\
KIC 09651065$^{e}$       &	273.8(3)	&	19.47768&1.931(2) ($Kp$) &O$-$C	&	1.7*(2)	&	2.6*(5)	&	7010*(140) & 3.8*(2)	&		     &		     &		    &		    &	HUB14, SHI15, MUS15	            \\
KIC 10080943$^{d9, e}$ A &15.3364(3)    &15.683330(3)&		         &SB2+EV&	2.0(1)	&	2.9(1)	&	7100(200)  &	3.81(3)	&	 1.9(1)	 &	2.1(2)	 &	40.76	&	0.96(1)	&	SCH15, KEE15, SCH16	            \\
KIC 10080943$^{d9, e}$ B &              &13.947586(2)&		         &	    &	1.9(1)	&	2.1(2)	&	7480(200)  &	4.1(1)	&	      	 &	     	 &	        &	       	&	                                \\
KIC 10990452$^{e}$	     &	122.11(36)	&  17.7237	 &5.654(3) ($Kp$)&O$-$C & 1.7*(3)	&	2.1*(4)	&	7622*(266) &	4.0*(4)	&		     &	         &		    &		    &	HUB14, SHI15, MUS15	            \\
KIC 11754974$^{e}$       &	343.27(34)	&16.3447449(3)&51.79(5) ($Kp$)&SB1+O$-$C&1.53*	&	1.76	&	  7256	   &	4.129	&		     &		     &		    &		    &	MUR13a, MUR13b	                \\
RY Lep	                 &	730	        &	4.4415	 &96.8(4) ($I$)	 &	SB1	&		    &		    &		       &		    &		     &		     &		    &		    &	LAN03, DER09	                \\
AN Lyn	                 &   6449(94)   &10.175583(2)&	94.0 ($V$) 	 &O$-$C	&	2.18*   &		    &	7800*	   &	3.8*	&		     &		     &		    &		    &	ROD97, ZHO02, HIN05,            \\
                &              &              &                &          &             &           &              &            &            &           &          &           &   LI10, PEN15                     \\
BE Lyn	                 &	3475	    &10.430844(2)&172.5(8) ($V$) &O$-$C	&		    &		    &	8700*	   &	3.8*	&		     &		     &		    &		    &	DER09, KIM12, PEN15	            \\
SZ Lyn$^{e}$	         &	1186(5)	    &	8.2965(6)&	 222 ($V$)	 &SB1+O$-$C	&		&		    &		       &		    &		     &		     &		    &		    &	ROD00, POU04, GAZ04, LI13	    \\
IK Peg$^{d10}$           &	21.722      &	22.7272	 &	 10 ($V$)	 &	SB1	&	1.69*	&		    &	7600*	   &		    &		     &		     &     		&		    &	WON93, WON94, 	                \\
                &              &              &                &          &             &           &              &            &            &           &          &           &   ROD00, DUC11                    \\
$\delta$ Ser$^{e}$       &	379119	    &	6.4227	 &	 22.5 ($V$)	 &	 Vis	&		&		    &		       &		    &		     &		     &		    &		    &	ROL87, ROD00, MAL12	            \\
$\theta^2$ Tau$^{d11, e}$ &	140.7282(9)	&	13.220	 &	4.8(1) ($v$) &	SB2	&	2.86(6)	&	4.4(1)	&	7800(170)  &  3.6(1)	&	2.16(2)	 &	 2.7(1)	 &	199(2)  &	0.754(2)&   BRE02, POR02, TOR11	            \\
$\rho$ Tau$^{d12}$	     &	488.5       &	14.925	 &	10 ($V$)	 &	SB1	&	1.6*	&		    &	7300*	   &		    &		     &		     &		    &		    &	HOR79, ANT98, DUC11	            \\
V479 Tau$^{d13, e}$ 	 &    85722     &	13.1980	 &	30 ($V$)	&	Vis	&		    &		    &		       &		    &		     &		     &		    &		    &   DIC67, MAL12, HAR13             \\
V777 Tau$^{d14, e}$	     &	  5200	    &	5.485	 &	6.0(7) ($V$) &	SB1	&		    &		    &	7300*	   &		    &		     &		     &		    &		    &	KRI95, ROD00, POU04	            \\
GW UMa	                 &	4821(256)   &4.92141(1)	 &	450 ($V$)	 &O$-$C	&	1.76*	&		    &	7600*	   &		    &	    	 &		     &		    &		    &	HIN05, WAN15	                \\
FM Vir$^{d15, e}$	        &	38.324	&	13.9119	&	20 ($V$)	&	SB2	&	2.05*	&		    &		       &		    &	1.9*	 &		     &		    &	0.927	    &	BER57, MIT79, BAR83, POU04	    \\
\hline
\end{tabular}}
\newline
*assumed, $^{e}$eccentric orbit, $^{d1}$4~CVn, $^{d2}$HD~197461, both components (referred as A and B) are pulsators, $^{d3}$HD~104288, $^{d4}$BD-04~1759, $^{d5}$ASAS~163118+1159.8, $^{d6}$HIP~96299, Hybrid, $^{d7}$HD~181469, quintuple system, $^{d8}$hybrid, $^{d9}$both components (referred as A and B) are hybrid pulsators, $^{d10}$HD~204188, HR~8210, $^{d11}$HD~28319, $^{d12}$86~Tau, HD~28910, HIC 21273, $^{d13}$HD~24550, $^{d14}$71~Tau, $^{d15}$32~Vir, d02~Vir\\
ANT98-\citet{ANT98}, BAA93-\citet{BAA93}, BAR83-\citet{BAR83}, BER57-\citet{BER57}, BRE02-\citet{BRE02}, BRE08-\citet{BRE08}, CAT10-\citet{CAT10}, DER09-\citet{DER09}, DIC67-\citet{DIC67}, DUC11-\citet{DUC11}, FRA95-\citet{FRA95}, FU99-\citet{FU99}, FU00-\citet{FU00}, FU03-\citet{FU03}, FU08-\citet{FU08}, GAZ04-\citet{GAZ04}, GRE13-\citet{GRE13}, HAR13-\citet{HAR13}, HAR14-\citet{HAR14}, HAR84-\citet{HAR84}, HEN01-\citet{HEN01}, HIN05-\citet{HIN05}, HOR79-\citet{HOR79}, HUB14-\citet{HUB14}, HUR07-\citet{HUR07}, KEE15-\citet{KEE15}, KHO07-\citet{KHO07}, KIM12-\citet{KIM12}, KRI95-\citet{KRI95}, LAN03-\citet{LAN03}, LEE12-\citet{LEE12}, LI10-\citet{LI10}, LI13-\citet{LI13}, LOP79-\citet{LOP79}, MAC14-\citet{MAC14}, MAL12-\citet{MAL12}, MAW11-\citet{MAW11}, MIL94-\citet{MIL94}, MIT79-\citet{MIT79}, MUR13a-\citet{MUR13a}, MUR13b-\citet{MUR13b}, MUS15-\citet{MUS15}, PEN15-\citet{PEN15}, POC00-\citet{POC00}, POR02-\citet{POR02}, POU04-\citet{POU04}, PRU11-\citet{PRU11}, REI76-\citet{REI76}, ROD97-\citet{ROD97}, ROD98a-\citet{ROD98AQR}, ROD00-\citet{ROD00}, ROL87-\citet{ROL87}, SCH14-\citet{SCH14}, SCH15-\citet{SCH15}, SCH16-\citet{SCH16}, SHI12-\citet{SHI12}, SHI15-\citet{SHI15}, STE12-\citet{STE12}, TOR11-\citet{TOR11}, WAN15-\citet{WAN15}, WON93-\citet{WON93}, YAN93-\citet{YAN93}, ZHO02-\citet{ZHO02}
\end{table}
\end{landscape}

\begin{landscape}

\begin{table}
\caption{List of binaries with unspecified $P_{\rm orb}$ containing a $\delta$~Sct component.}
\label{tab:tab3}
\scalebox{0.94}{
\begin{tabular}{lcccl lcccl}
\hline
        Name	    &$f_{\rm dom} $	&	$A$	    &	ToV &	     References\hspace{3cm}        &     Name	           &$f_{\rm dom} $	&	$A$	       &	ToV   &	References            \\
	                &(cycle~d$^{-1}$)&(mmag)	&		&	                                   &                       &(cycle~d$^{-1}$)& (mmag)       &		  &                       \\
\hline
GN And$^{1}$	    &  14.4282  & 1.52(7) ($V$) &	Vis	&	GAR85, TUR93,                      &   HD 050870$^{19}$	   &	17.1617	    &12.96($CoR$)  &	SB2	  &	MAN12                 \\
                    &           &               &		&	ROD98b, DALL02                     &   HD 064813$^{5}$	   &  20.3100(1)	&2.9(1) ($V$)  &	Vis   &	KIMD10, HAR13         \\
V361 And$^{2}$ 	    &	7.4963	&	18.27 ($V$)	&	Vis	&	FAB02, LI07	                       &	HD 189503$^{20}$   &	 5.1918	    &  110 ($V$)   &	Vis	  &	MAS01, POJ02	      \\
V419 And$^{3}$ 	    &	19.4090	&	6.96 ($V$)	&	Vis	&	MAR99, SMA11	                   &	HD 218994	       &	     24.7	&2.9(2) ($B$)  &	Vis	  &	GON07, GON08, KUR08	  \\
V1208 Aql$^{4}$ 	&6.68072(6)	&	18.1 ($V$)	&	Vis	&	BRE76, PRO81,                      &	HD 292962$^{21}$   &	9.18898	    &	200 ($V$)  &	Vis	  &	MAS01, POJ02	      \\
                    &           &               &		&	WEH95, DALL02                      &	LM Hya$^{22}$ 	   &	26.323(7)	& 2.9(1) ($B$) &	Vis	  &	KUR84, MAS01	      \\
$\alpha$~Aql$^{5}$	&	15.768	&	 	        &	Vis	&	BUZ05, MAS01                       &	BN Hyi$^{23}$ 	   &	 15.4(2)	&	20 ($V$)   &	Vis	  &	MOO83, FAB02	      \\
OX Aur$^{6}$ 	    &	6.8283	&	6.1 ($v$)	&	Vis	&	ZHI00b, MAS01	                   &	KIC 04840675$^{5}$ &  22.06872(1)   &	1.3715(9)  &	 SB	  &	BAL12	              \\
BD+18 4988$^{7}$	&5.779545   &	130 ($V$)	&	Vis	&	FAB02, POJ02	                   &	CC Lyn	           &	5.6402(4)	&24.5(1) ($W$) & Vis+SB1  &	PRI09, CON10	      \\
BD$-$05 3329$^{8}$  &	6.43165	&	200 ($V$)	&	Vis	&	MAS01, POJ02	                   &	XZ Men$^{24}$ 	   &	9.231	    &	4.7 ($B$)  &	Vis	  &	KUR80, MAS01	      \\
$\gamma$ Boo$^{9}$ 	&	21.28   &	 	        &	Vis	&	AUV79, MAS01,                      &	V752 Mon$^{5}$	   &	4.3202	    &	24.5 ($Hp$)&	Vis	  & PRI06, RUC09,         \\
                    &           &               &		&	VEN07, DALL02                      &	                   &	            &              &		  &	RUC07, DUB11	      \\
$\iota$ Boo	        &37.750(1)	&	3.5 ($V$)	&	Vis	&	KIS95, KIS99, MAS01	               &	LW Mus$^{25}$      &	6.5088	    &	140 ($V$)  &	Vis	  &	MAS01, POJ02	      \\
X Cae$^{10}$ 	    &	7.394	&	37.2 ($V$)	&	Vis	&	MAN00, FAB02	                   &	CP Oct$^{26}$	   &6.67766(4)	    &	25 ($Hp$)  &	Vis	  &	ESA97, MAS01, GON08	  \\
V527 Car$^{11}$ 	&4.680412	&	29(1) ($V$)	&	Vis	&	KOE99, FAB02	                   &	V2542 Oph$^{27}$   &	13.2624(2)	&11.4(7) ($V+y$)&	Vis	  &	KAY00, MAS01	      \\
$\beta$ Cas	        &9.897396(5)&15.4(7) ($V$)	&	Vis	&	RIB94, MAS01, GUI13	               &	$\rho$ Pup         &   7.098168(1)  &	95 ($V$)   &	Vis	  & PON63, MAS01, MOO09	  \\
V1063 Cen$^{12}$ 	&	20.0557	&	7.3 ($B$)	&	Vis	&	FAB02, MAR02	                   &	$\delta$ Sct 	   &	5.16077(1)	&	200 ($V$)  &	Vis	  &	STE38, MOO82, MAS01	  \\
V377 Cep 	        &13.705(1)	&	18 ($R$)	&	Vis	&	KUS95, FAB02	                   &	RX Sex$^{28}$ 	   &	12.5156	    &	10 ($V$)   &	Vis	  &	JER72, ABT81	      \\
HS Eri	            &	4.98902	&	18.3 ($V$)	&	Vis	&	HOR00, KOE02	                   &TYC 0313-337-1$^{29}$  &	 6.6892	    &	210 ($V$)  &	Vis	  &	MAS01, POJ02	      \\
IU Eri$^{13}$	    &14.51837(6)&11.8(6) ($V$)	&	Vis	&	MAS01, HANS02	                   &TYC 5827-482-1$^{30}$  &	 17.7120	&	120 ($V$)  &	Vis	  &	MAS01, POJ02	      \\
TY For$^{14}$	    & 10.2942	&	15 ($V$)	&	Vis	&	EGG79, LAM90, FAB02	               &	DP UMa$^{31}$	   &	    25	    &	20 ($V$)   &	Vis	  &	KUR78, KUR79, MAS01	  \\
VV For$^{15}$	    &	17.3310	&10.1(2) ($v$)	&	Vis	&	STE89, MAS01	                   &	$\upsilon$ UMa	   &	 7.54(2)	&	71(8) ($V$)&	Vis	  &	DWO73, MAS01	      \\
PV Gem	            &	5.3173	&	50 ($V$)	&	Vis	&	NIC10, HAR13	                   &	FZ Vel 	           &	15.3846	    &	15 ($V$)   &	Vis	  &	EGG70, HAR13	      \\
DQ Gru$^{16}$	    &	4.679	&	15.8 ($B$)	&Vis+SB2&	LAM92, TSV93,                      &	V540 Vel$^{32}$    &	5.9192	    &	46 ($V$)   &	Vis	  &	MAS01, KHR10	      \\
                    &           &               &		&	LAM99, LAM00                       &	FG Vir	           &	12.716	    &	22.1 ($Y$) &	Vis	  &	ABT81, LEN08, MAS01   \\
HD 011667$^{17}$    &10.9888	&	290 ($V$)	&	Vis	&	MAS01, POJ02	                   &	GG Vir$^{33}$	   &	17.8571	    &	13 ($B$)   &	Vis	  &	BAR75, STE77, MAS01	  \\
HD 017421$^{18}$    &	12.4510	&	80 ($V$)	&	Vis	&	FAB02, POJ02	                   &	OV Vir$^{34}$	   &	31.3548(2)	& 4.9(1) ($b$) &	Vis	  &	RODR01, MAS01	      \\
HD 041140             &  25.827(1)&  4.7(2) ($V$) &   Vis &	FAB02, KIMD10                      &		           &	    &	 &	  &\\

\hline
\end{tabular}}
\newline
$^{1}$28~And, $^{2}$HD~6859, $^{3}$HD~13079, $^{4}$28~Aql, HR~7331, $^{5}$multiple system, $^{6}$59~Aur, $^{7}$ASAS~J222629+1924.8, $^{8}$ASAS~J114026-0606.2, $^{9}$triple system, $^{10}$$\gamma^2$~Cae, $^{11}$HD~95321, $^{12}$HD~129052, $^{13}$HD~27093, $^{14}$HIP~11644, HR~733, HD~15634 $^{15}$HD~17978, $^{16}$HD~220392, $^{17}$ASAS~J015307-5056.5, $^{18}$ASAS~J024816+2213.4, $^{19}$BD$-$03~1642, $^{20}$ASAS~J200221-4554.0, $^{21}$ASAS~J065937-0047.0, $^{22}$2~Hya, HR~3321, $^{23}$HR~981, $^{24}$HD~31908, $^{25}$ASAS~J133845-7011.2, $^{26}$HD~21190, $^{27}$HD~152569, $^{28}$HD~90386, $^{29}$ASAS~J133737+0715.9, $^{30}$ASAS~J231240-1305.8, $^{31}$67~UMa, HR~4594, $^{32}$HD~298732, $^{33}$27~Vir, $^{34}$HD~129231\\
ABT81-\citet{ABT81}, AUV79-\citet{AUV79}, BAL12-\citet{BAL12}, BAR75-\citet{BAR75}, BRE76-\citet{BRE76}, BUZ05-\citet{BUZ05}, CON10-\citet{CON10}, DALL02-\citet{DALL02}, DUB11-\citet{DUB11}, DWO73-\citet{DWO73}, EGG70-\citet{EGG70}, EGG79-\citet{EGG79}, ESA97-\citet{ESA97}, FAB02-\citet{FAB02}, GAR85-\citet{GAR85}, GON07-\citet{GON07}, GON08-\citet{GON08}, GUI13-\citet{GUI13}, HANS02-\citet{HANS02}, HAR13-\citet{HAR13}, HOR00-\citet{HOR00}, JER72-\citet{JER72}, KAY00-\citet{KAY00}, KHR10-\citet{KHR10}, KIMD10-\citet{KIMD10}, KIS95-\citet{KIS95}, KIS99-\citet{KIS99}, KOE99-\citet{KOE99}, KOE02-\citet{KOE02}, KUR78-\citet{KUR78}, KUR79-\citet{KUR79}, KUR80-\citet{KUR80}, KUR84-\citet{KUR84}, KUR08-\citet{KUR08}, KUR15-\citet{KUR15}, KUS95-\citet{KUS95}, LAM90-\citet{LAM90}, LAM92-\citet{LAM92}, LAM99-\citet{LAM99}, LAM00-\citet{LAM00}, LEN08-\citet{LEN08}, LI07-\citet{LI07}, MAN00-\citet{MAN00}, MAN12-\citet{MAN12}, MAR99-\citet{MAR99}, MAR02-\citet{MAR02}, MAS01-\citet{MAS01}, MOO82-\citet{MOO82}, MOO83-\citet{MOO83}, MOO09-\citet{MOO09}, NIC10-\citet{NIC10}, PON63-\citet{PON63}, POJ02-\citet{POJ02}, PRI06-\citet{PRI06}, PRI09-\citet{PRI09}, PRO81-\citet{PRO81}, RIB94-\citet{RIB94}, ROD98b-\citet{ROD98AND}, RODR01-\citet{RODR01}, RUC07-\citet{RUC07}, SMA11-\citet{SMA11}, STE77-\citet{STE77}, STE89-\citet{STE89}, STE38-\citet{STE38}, TSV93-\citet{TSV93}, TUR93-\citet{TUR93}, VEN07-\citet{VEN07}, WEH95-\citet{WEH95}, ZHI00b-\citet{ZHI00Aur}
\end{table}
\end{landscape}

\begin{landscape}
\begin{table}
\caption{Ambiguous cases}
\label{tab:tab4}
\scalebox{0.85}{
\begin{tabular}{lll lll}
\hline
Name		&	Comments & References\hspace{3.8cm} & Name		&	Comments & References	\\
\hline
\multicolumn{6}{c}{Confirmed binary systems but ambiguous for pulsations type/existence}\\
\hline
AD Ari	        &	EV, but weak proof for pulsations 	        &	HANS02	            &	KIC 06963171	&	Unclear type of pulsations  	&	BRA15 	\\
VW Ari	        &	Visual triple but probably of $\lambda$ Boo type 	&	MUR15a	    &	KIC 08553788	&	Unclear type of pulsations 	&	GIE12 	\\
ASAS 073904$-$6037.2$^{1}$	&	Weak proof for pulsation type	&	 PIG07	            &	KIC 08553788	&	Unclear type of pulsations  	&	GIE12	\\
ASAS 110615$-$4224.6$^{2}$	&	Weak proof for pulsation type	&	 PIG07	            &	KIC 08895509	&	Unclear type of pulsations  	&	BRA15 	\\
ASAS 234520$-$3100.5$^{3}$	&	Weak proof for pulsation type	&	 PIG07	            &	KIC 11819135	&	Unclear type of pulsations  	&	BRA15 	\\
NR CMa	        &	Vis. but weak proof for pulsations 	&	ESA97, MAS01	            &	KIC 11867071	&	Unclear type of pulsations  	&	BRA15 	\\
MM Cas 	        &	Weak proof for pulsations 	&	 CHA83, CHA92	                    &  $\beta$ Leo	&	Vis. but unclear type of pulsations 	&	BAR81, MKR98, MAS01	\\
V377 Cas	    &	Vis. but weak proof for pulsations 	&	LOW89, FAB00	            &	DT Lup	&	Weak proof for pulsations 	&	MID11	\\
$\psi$ Cen 	    &	Unclear presence and/or type of pulsations  	&	BRU06, MAN10    &	$\alpha$ Lyr 	&	Vis. but weak proof for pulsations 	&	BOH12, BOH15	\\
V360 Cep	    &	Vis. but weak proof for pulsations 	&	MAR82, FAB00	            &	EY Ori 	&	Weak proof for pulsations 	&	ZAK98, SMA14, TUR14	\\
UU Com$^{4}$	&	Probably it is of $\alpha^{2}$ CVn type 	&	SAN89, SAV96, LIP08 &	FO Ori 	&	Weak proof for pulsations 	&	TUR14	\\
$\beta$ CrB 	&	Probably it is of $\alpha^{2}$ CVn type 	&	 HAT04, KUR07, BRU10&	HT Peg	&	Vis. but weak proof for pulsations 	&	CON74, MAS01	\\
$\gamma$ CrB 	&	Vis. but unclear type of pulsations 	&	VET81, LEH97, MAL12	    &	NN Peg	&	Vis. but weak proof for pulsations 	&	FAB02	\\
$\tau$ Cyg	    &	Vis. but unclear type of pulsations 	&	MKR95, FAB02	        &	V368 Pup	&	Vis. but weak proof for pulsations 	&	ESA97, MAS01	\\
AI Dra	        &	Weak proof for pulsations 	&	 NAR02b, KIS02, LAZ04 	            &	V1060 Sco	&	Vis. but weak proof for pulsations 	&	ESA97, MAS01	\\
CC Gru	        &	Vis. but weak proof for pulsations 	&	ESA97, MAS01	            &	V1072 Sco	&	Vis. but weak proof for pulsations 	&	ESA97, MAS01	\\
HD 073712	    &	SB2  but doubtful for pulsations 	&	MIC95, DUC11	            &	V1241 Tau$^{6}$	&	The pulsating behaviour is questionable 	&	 ARE04, PIG07, ULA14 	\\
HD 120635	    &	Vis. but weak proof for pulsations 	&	MAS01, RIC12	            &	TYC 7817-75-1	&	Vis. but weak proof for pulsations 	&	MAS01, RIC12	\\
V979 Her 	    &	Vis. but unclear spectral type ($\beta$~Cep?)	&	MAS01, POJ02	&	X Tri	&	Weak proof for pulsations 	&	 KIM03, LIA10, TUR14	\\
V994 Her$^{5}$	&	Quintuple system with 2 E 	&	 DAL05, LEE08, ZAS13 	            &  RU UMi	&	Weak proof for pulsations 	&	BEL93, MAN01, ZHU06	\\
KIC 04470124	&	Unclear type of pulsations 	&	BRA15  	                            &	GZ Vir 	&	Triple system but weak proof for pulsations 	&	PEN81, ROD88, TOK10	\\
KIC 05615815	&	Unclear type of pulsations  	&	BRA15 	                        &	HX Vir$^{7}$	&	Vis. but weak proof for pulsations 	&	STE86, MAS01	\\
KIC 05878081	&	Unclear type of pulsations  	&	BRA15 	                        &	$\theta$ Vir$^{8}$	&	Weak proof for pulsations 	&	BEA77, ADE97	\\
\hline
\multicolumn{6}{c}{Confirmed $\delta$~Sct stars but ambiguous for binarity}\\
\hline
V853 Cep	       &	Weak proof for visual binarity 	&	VAN86	                    &	V1366 Ori$^{11}$	&	Weak proof for binarity 	&	SOL03, AMA04, CAS13	\\
KIC 05988140$^{9}$ &Unclear for binarity. Hybrid $\gamma$~Dor$-\delta$~Sct star&LAM13   &	DL UMa	&	Weak proof for binarity 	&	ROD00, NIC10 	\\
V474 Mon$^{10}$	   &	Weak proof for binarity 	&	 DES74, BAL01, AMA07 	        &		&		&		\\
\hline
\multicolumn{6}{c}{Ambiguous for both binarity and pulsations type/existence}\\
\hline
HV Eri	           &	It might be of W UMa type 	&	PRI09	                        &	o Ser	&	Weak proof for SB and for pulsations 	&	VAL72, DWO75	\\
BR Hyi	           &	It might be of W UMa type 	&	POJ02	                        &	V353 Vel	&	It might be of W UMa type 	&	PRI03, PRI06	\\
V1359 Ori	       &	It might be of W UMa type 	&	PRI09	                        &		&		&		\\
\hline
\end{tabular}}
\newline
$^{1}$CPD$-60\degr871$, $^{2}$CPD$-41\degr5106$, $^{3}$CPD$-31\degr6830$, $^{4}$21~Com, HD~108945, $^{5}$HD~170314, $^{6}$WX~Eri, $^{7}$73~Vir, $^{8}$51~Vir, $^{9}$HD~188774, $^{10}$1~Mon, HD~40535, $^{11}$HD~34282\\
ADE97-\citet{ADE97}, AMA04-\citet{AMA04}, AMA07-\citet{AMA07}, ARE04-\citet{ARE04}, BAL01-\citet{BAL01}, BAR81-\citet{BAR81}, BEA77-\citet{BEA77}, BEL93-\citet{BEL93}, BOH12-\citet{BOH12}, BOH15-\citet{BOH15}, BRA15-\citet{BRA15}, BRU06-\citet{BRU06}, BRU10-\citet{BRU10}, CAS13-\citet{CAS13}, CHA83-\citet{CHA83}, CHA92-\citet{CHA92}, CON74-\citet{CON74}, DAL05-\citet{DAL05}, DES74-\citet{DES74}, DUC11-\citet{DUC11}, DWO75-\citet{DWO75}, ESA97-\citet{ESA97}, FAB00-\citet{FAB00}, FAB02-\citet{FAB02}, GIE12-\citet{GIE12}, HANS02-\citet{HANS02}, HAT04-\citet{HAT04}, KIM03-\citet{KIM03}, KIS02-\citet{KIS02}, KUR07-\citet{KUR07}, LAM13-\citet{LAM13}, LAZ04-\citet{LAZ04}, LEE08-\citet{LEE08}, LEH97-\citet{LEH97}, LIA10-\citet{LIA10}, LIP08-\citet{LIP08}, LOW89-\citet{LOW89}, MAL12-\citet{MAL12}, MAN01-\citet{MAN01}, MAN10-\citet{MAN10}, MAR82-\citet{MAR82}, MAS01-\citet{MAS01}, MIC95-\citet{MIC95}, MID11-\citet{MID11}, MKR95-\citet{MKR95}, MKR98-\citet{MKR98}, MUR15a-\citet{MUR15a}, NAR02b-\citet{NAR02b}, NIC10-\citet{NIC10}, PEN81-\citet{PEN81}, PIG07-\citet{PIG07}, POJ02-\citet{POJ02}, PRI03-\citet{PRI03}, PRI06-\citet{PRI06}, PRI09-\citet{PRI09}, RIC12-\citet{RIC12}, ROD88-\citet{ROD88}, ROD00-\citet{ROD00}, SAN89-\citet{SAN89}, SAV96-\citet{SAV96}, SMA14-\citet{SMA14}, SOL03-\citet{SOL03}, STE86-\citet{STE86}, TOK10-\citet{TOK10}, TUR14-\citet{TUR14}, ULA14-\citet{ULA14}, VAL72-\citet{VAL72}, VAN86-\citet{VAN86}, VET81-\citet{VET81}, ZAK98-\citet{ZAK98}, ZAS13-\citet{ZAS13}, ZHU06-\citet{ZHU06}
\end{table}
\end{landscape}

\begin{landscape}
\begin{table}
\caption{Rejected cases}
\label{tab:tab5}
\begin{tabular}{lll lll}
\hline
Name		 &	Comments & References\hspace{3.5cm}  & Name		&	Comments & References	\\
\hline
CC And	     &	No proof for binarity 	&	EKM08	                      &	GSC 3382-0957	&	E or an hybrid $\gamma$~Dor$-\delta$~Sct star 	&	 ZHA12	\\
ET And	     &	It is of $\alpha^{2}$ CVn type (B9p(Si) star) &	 WEI98	  &	HD 086731	&	No relevant references exist 	&		\\
GP And	     &	No proof for binarity 	&	ZHO11	                      &	HD 094529	&	No proof for pulsations 	&	 PIG07	\\
V529 And$^{1}$ 	 &	No proof for binarity 	&	HEN05	                  &	HD 188164	&	It is of $\lambda$ Boo type 	&	MUR15a 	\\
DV Aqr	     &	No proof for pulsations 	&	POL10	                  &	CN Hyi	&	It is of W UMa type	&	RUC06	\\
$\zeta$ Aur	 &	Prototype of $\zeta$ Aur type systems 	&	EAT08 	      &	KIC 09204718$^{5}$	&	E or an hybrid $\gamma$~Dor$-\delta$~Sct star	&	TUR13	\\
BR Cnc	     &	Member of OC. Not Binary*	&	ZHOU01	                  &	UX Mon	&	It is of W Ser type 	&	SUD11	\\
EX Cnc	     &	Member of OC. Not Binary*	&	ZHO01b	                  &	DE Oct	&	It is of W UMa type 	&	PRI06	\\
HI Cnc	     &	Member of OC. Not Binary*	&	ZEJ12	                  &	BQ Phe	&	It is of W UMa type 	&	POU04	\\
VZ CVn 	     &	It is of $\gamma$~Dor type 	&	LAT12	                  &	V856 Sco	&	It is of UX Ori type	&	NAT97, BEN11	\\
AZ CMi 	     &	No proof for binarity 	&	FED91	                      &	$\upsilon$ Tau	&	Member of OC. Not Binary*	&	BOS83	\\
V410 Car	 &	Member of OC. Not Binary*	&	ANT86	                  &	V480 Tau	&	Member of OC. Not Binary*	&	KIP78	\\
RX Cas 	     &	It is of W Ser type 	&	TOD89, TAR97, PUS07 	      &	V534 Tau	&	Member of OC. Not Binary*	&	LI04	\\
V965 Cep 	 &	No proof for binarity 	&	SOK09	                      &	V647 Tau	&	Member of OC. Not Binary*	&	LIU99	\\
AV Cet 	     &	No proof for binarity 	&	MAN05 	                      &	V650 Tau	&	Member of OC. Not Binary*	&	FOX11	\\
AI Com$^{2}$ &	It is of $\alpha^{2}$ CVn type. Vis. + SB	&ADE81, SAV96 &	V1128 Tau	&	It is of W UMa type 	&	CAL14	\\
TZ CrB$^{3}$ &	It is of RS CVn type 	&	 FRA97 	                      &	V1178 Tau	&	Member of OC. Not Binary*	&	ARE05	\\
V1644 Cyg$^{4}$&	It is of $\lambda$ Boo type 	&	CAS09	          &	V1229 Tau$^{6}$	&	No proof for pulsations 	&	GRO07	\\
V2448 Cyg	 &	Member of OC. Not Binary*	&	FRE01	                  &	TYC 1845-2532-1	&	Member of OC. Not Binary*	&	POJ02, ZEJ12	\\
ZZ Cyg	     &	No proof for pulsations 	&	 YAN15	                  &	GU Vel	&	No proof for binarity 	&	EGG08	\\
V345 Gem	 &	It is of W UMa type 	&	PRI06	                      &	V377 Vul$^{7}$	&	It is of SPB type 	&	 HUB91, DEC07, WAL12	\\
\hline
\end{tabular}
\newline
SPB=Slowly pulsating B-type star, OC=Open cluster.\\
*The star belongs to an OC but it is not member of a binary system.\\
$^{1}$HD~8801, $^{2}$17~Com, $^{3}$$\sigma_{2}$~CrB, $^{4}$29~Cyg, $^{5}$HD~176843, $^{6}$HD~23642, $^{7}$3~Vul, HD~182255   \\
ADE81-\citet{ADE81}, ANT86-\citet{ANT86}, ARE05-\citet{ARE05}, BEN11-\citet{BEN11}, BOS83-\citet{BOS83}, CAL14-\citet{CAL14}, CAS09-\citet{CAS09}, DEC07-\citet{DEC07}, EAT08-\citet{EAT08}, EGG08-\citet{EGG08}, EKM08-\citet{EKM08}, FED91-\citet{FED91}, FOX11-\citet{FOX11}, FRA97-\citet{FRA97}, FRE01-\citet{FRE01}, GRO07-\citet{GRO07}, GUI13-\citet{GUI13}, HEN05-\citet{HEN05}, HUB91-\citet{HUB91}, KIP78-\citet{KIP78},   LAT12-\citet{LAT12}, LEH03-\citet{LEH03}, LI04-\citet{LI04}, LIU99-\citet{LIU99}, MAN05-\citet{MAN05}, MUR15a-\citet{MUR15a}, NAT97-\citet{NAT97}, PIG07-\citet{PIG07}, POJ02-\citet{POJ02}, POL10-\citet{POL10}, POU04-\citet{POU04}, PRI06-\citet{PRI06}, PUS07-\citet{PUS07}, RUC06-\citet{RUC06}, SAV96-\citet{SAV96}, SOK09-\citet{SOK09}, SUD11-\citet{SUD11}, TAR97-\citet{TAR97}, TOD89-\citet{TOD89}, TUR13-\citet{TUR13},      WAL12-\citet{WAL12}, WEI98-\citet{WEI98}, YAN15-\citet{YAN15}, ZEJ12-\citet{ZEJ12}, ZHA12-\citet{ZHA12}, ZHO01b-\citet{ZHO01b}, ZHOU01-\citet{ZHOU01}, ZHO11-\citet{ZHO11}
\end{table}
\end{landscape}

\begin{figure}
\includegraphics[width=\columnwidth]{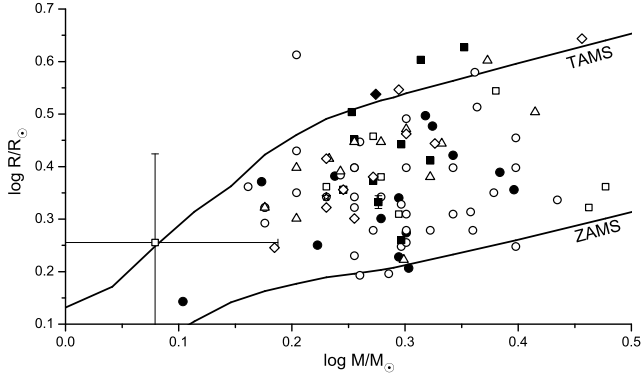}
\caption{The distribution of binary $\delta$~Sct stars within the $M-R$ diagram. Filled symbols denote the stars with very accurate measured parameters, while the open ones those with less accurate parameters. Circles, squares, and triangles symbols correspond to the stars that belong to semi-detached, detached, and unclassified systems with $P_{\rm orb}<13$~days, respectively. Diamond and star symbols correspond to those that belong to detached and unclassified systems with $P_{\rm orb}>13$~days, respectively. To avoid any clutter and make the figures easier to read, only the largest and the smallest error bars that delimit the range of uncertainties are plotted.}
    \label{fig:statMR}
\end{figure}

In Fig.~\ref{fig:statMR} the location of the binary $\delta$~Sct stars within the Mass-Radius ($M-R$) diagram is shown. Zero and Terminal Age (ZAMS and TAMS, respectively) Main Sequence lines (for solar metallicity composition, i.e. $Z$=0.019) were taken from \citet{GIR00}. Stars with accurately calculated parameters are considered those which are double-line spectroscopic eclipsing binaries, while all of the parameters of the rest of the systems are considered to be of lower accuracy. In this plot 84 $\delta$~Sct stars that belong to binaries with $P_{\rm orb}<13$~days (for sample details see Fig.~\ref{fig:stat2}) and 12 cases that belong to systems with $P_{\rm orb}>13$~days are shown.

\begin{figure}
\includegraphics[width=\columnwidth]{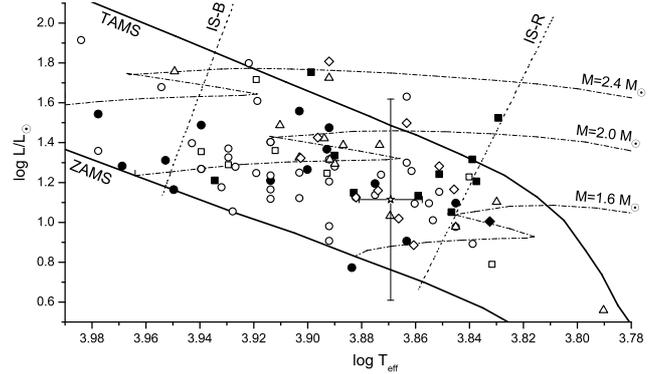}
\caption{The distribution of binary $\delta$~Sct stars within the $H-R$ diagram. Dotted lines represent the boundaries of the instability strip (Blue and Red edges) and the dash-dotted lines the evolutionary tracks for three different initial mass values. Symbols, lines, and error bars have the same meaning as in Fig.~\ref{fig:statMR}.}
    \label{fig:statHR}
\end{figure}

The distribution of the binary $\delta$~Sct stars is plotted within the Hertzsprung-Russell ($H-R$) in Fig.~\ref{fig:statHR}. ZAMS and TAMS lines as well as the evolutionary tracks for three different initial mass values and for solar metallicity composition were taken from \citet{GIR00}. The boundaries (B=Blue, R=Red) of the instability strip (IS) were taken from \citet{SOY366}. The sample is the same with that of Fig.~\ref{fig:statMR}, except for one case (V577~Oph) whose temperature has not been found yet.

\begin{figure}
\includegraphics[width=\columnwidth]{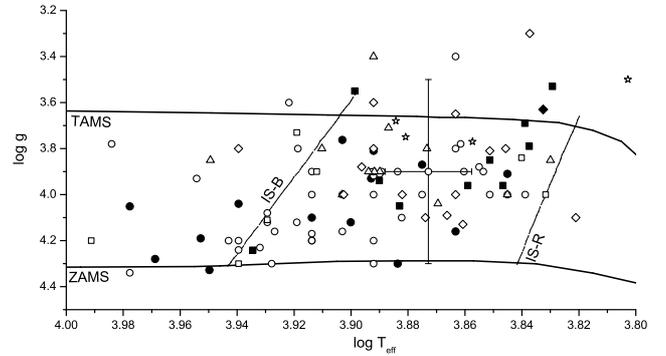}
\caption{The distribution of binary $\delta$~Sct stars within the $T_{\rm eff}-\log g$ diagram. Symbols, lines, and error bars have the same meaning as in Figs.~\ref{fig:statMR} and \ref{fig:statHR}.}
    \label{fig:statgT}
\end{figure}

The locations of the binary $\delta$~Sct stars within the $T_{\rm eff}-\log g$ diagram are given in Fig.~\ref{fig:statgT}. The sample consists of 87 $\delta$~Sct stars in binaries with $P_{\rm orb}<13$~days and 20 more stars in systems with $P_{\rm orb}>13$~days. The instability strip lines were taken from \citet{MUR15b} and correspond to modes with $l\leq2$, while ZAMS and TAMS lines, likely in Figs \ref{fig:statMR}-\ref{fig:statHR}, were taken from \citet{GIR00}.

\begin{figure}
\includegraphics[width=\columnwidth]{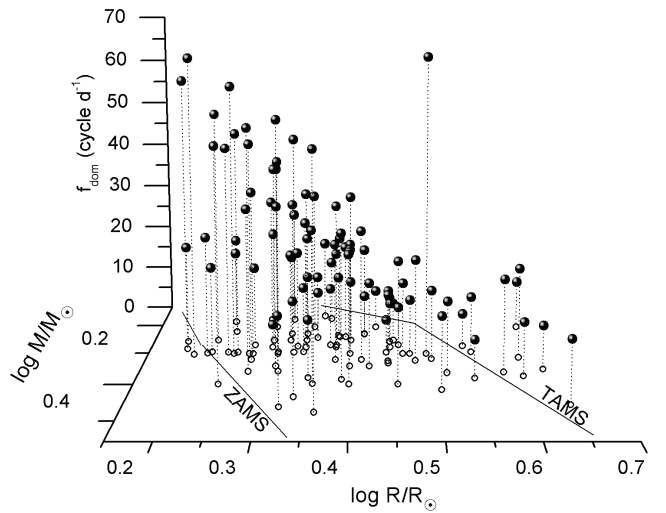}
\caption{The distribution of binary $\delta$~Sct stars according to their $f_{\rm dom}$ over the $M-R$ diagram. Dash lines indicate the projection of each point on the $M-R$ diagram.}
    \label{fig:stat3D}
\end{figure}

Fig.~\ref{fig:stat3D} contains the distribution of binary $\delta$~Sct stars according to their $f_{\rm dom}$ over the $M-R$ diagram. The sample is the same as in Fig.~\ref{fig:statMR}. Horizontal plane corresponds to the $M-R$ diagram, while the vertical axis to the $f_{\rm dom}$.

\section{Correlations}
\label{sec:s4}

Following previous works \citep[e.g.][]{SOY366, LIA12, ZHA13} correlations between fundamental stellar parameters are presented here for the systems with $P_{\rm orb}<13$~days (i.e. the first subgroup of Fig.~\ref{fig:statPP}).

\begin{figure}
\includegraphics[width=7.7cm]{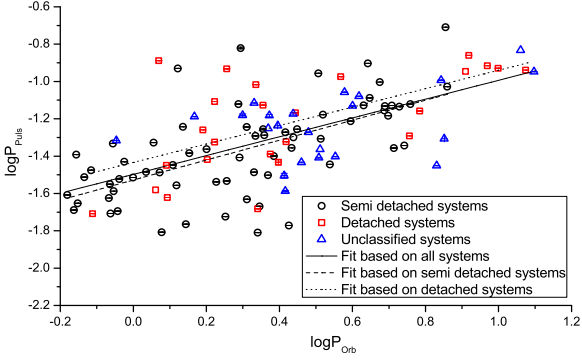}
\caption{The correlation between $\log P_{\rm orb}-\log P_{\rm pul}$ of binary $\delta$~Sct stars in systems with $P_{\rm orb}<13$~days. Three different linear fits (solid, dashed, and dashed-dotted lines) on different data samples (all, semi detached, and detached systems) are shown.}
    \label{fig:corlogPP}
\end{figure}

\begin{figure}
\includegraphics[width=7.7cm]{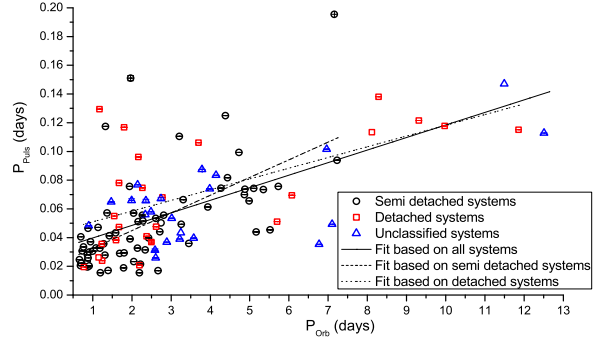}
\caption{The same as in Fig.~\ref{fig:corlogPP}, but for the decimal $P_{\rm orb}- P_{\rm pul}$ values.}
    \label{fig:corPP}
\end{figure}

The first correlation concerns the connection between $P_{\rm orb}-P_{\rm pul}$, which shows the dependence of the pulsations of the $\delta$~Sct star on the orbital period value of the system. In Figs~\ref{fig:corlogPP} and \ref{fig:corPP} the linear fits using logarithmic (Fig.~\ref{fig:corlogPP}) and decimal (Fig.~\ref{fig:corPP}) values of three different data samples are shown. The first fit was made on all data points (114), the second one on points that correspond to semi-detached systems (66), and the third one on those corresponding to detached systems (25). Plots and correlations are made for both logarithmic and decimal values so that can be directly compared with those of previous studies \citep{SOY366, LIA12, ZHA13}. The correlations between $P_{\rm orb}- P_{\rm pul}$ for each sample along with the respective correlation coefficient $r$ are presented below:

for $\delta$~Sct stars in \textsl{Semi Detached} binaries:
\begin{equation}
\log P_{\rm pul}= -1.53(3)+0.54(8)~\log P_{\rm orb},~with~r=0.62
    \label{cor:logPPSD}
\end{equation}
\begin{equation}
P_{\rm pul}= 0.020(6)+0.012(2)~P_{\rm orb}, with~r=0.61
    \label{cor:PPSD}
\end{equation}

for $\delta$~Sct stars in \textsl{Detached} binaries:
\begin{equation}
\log P_{\rm pul}= -1.43(7)+0.49(13)~\log P_{\rm orb},~with~r=0.60
    \label{cor:logPPD}
\end{equation}
\begin{equation}
P_{\rm pul}= 0.043(9)+0.007(2)~P_{\rm orb},~with~r=0.61
    \label{cor:PPD}
\end{equation}

for $\delta$~Sct stars in \textsl{all} known close binaries (Detached, Semidetached, unclassified):
\begin{equation}
\log P_{\rm pul}= -1.50(3)+0.50(6)~\log P_{\rm orb},~with~r=0.62
    \label{cor:logPPall}
\end{equation}
\begin{equation}
P_{\rm pul}= 0.031(4)+0.009(1)~P_{\rm orb},~with~r=0.62
    \label{cor:PPall}
\end{equation}

\begin{figure}
\includegraphics[width=7.7cm]{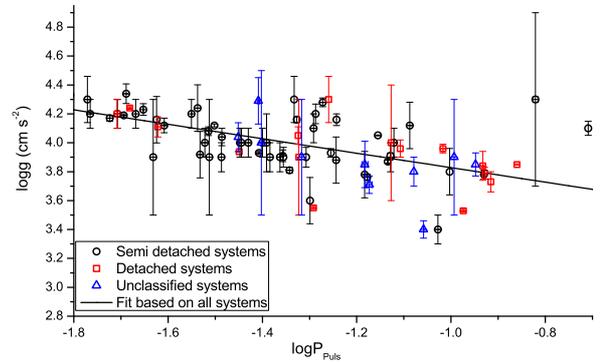}
\caption{The correlation between $\log P_{\rm pul}-\log g$ of binary $\delta$~Sct stars in systems with $P_{\rm orb}<13$~days.}
    \label{fig:corgP}
\end{figure}

The second correlation shows the dependence between the pulsation periods of the binary $\delta$~Sct stars and their evolutionary stages. Fig.~\ref{fig:corgP} shows the fit on the $\log g-\log P_{\rm pul}$ values of the binary $\delta$~Sct stars. It should be noted that the $\log g$ values of the pulsators, when not directly given in the literature, were calculated using the following formula:
\begin{equation}
\log g_{\rm pul}= 4.438 + \log M_{\rm pul} - 2\log R_{\rm pul}
\end{equation}
where 4.438 is the $\log g$ value of the Sun (in cm~s$^{-2}$), while the mass ($M_{\rm pul}$) and the radius ($R_{\rm pul}$) of the star are given in solar units.

The errors of $\log g$ values for 21 out of 82 binary $\delta$~Sct stars are not given in the literature. Therefore, their error values were assigned the average values of the errors of the stars of other systems whose absolute parameters were calculated based on the same available information. The reliability and the accuracy of the $\log g$ value and its error are directly connected to the available information for their derivation (e.g. RVs+light curves model, or light curves model only). In particular, we found 0.05 as an average $\log g_{\rm pul}$ error value for SB2+E, 0.16 for E, 0.06 for SB1, 0.1 for SB2+EV, and 0.07 for SB1+E systems. The whole sample consists of 82 stars, whose details are given in Table~\ref{tab:tab1} and their demographics are shown in Fig.~\ref{fig:stat2}. Based on the above assumptions, the following correlation is derived:

\begin{equation}
\log g= 3.33(1)-0.50(1)~\log P_{\rm pul},~with~r=0.68
    \label{cor:gP}
\end{equation}

\begin{figure}
\includegraphics[width=\columnwidth]{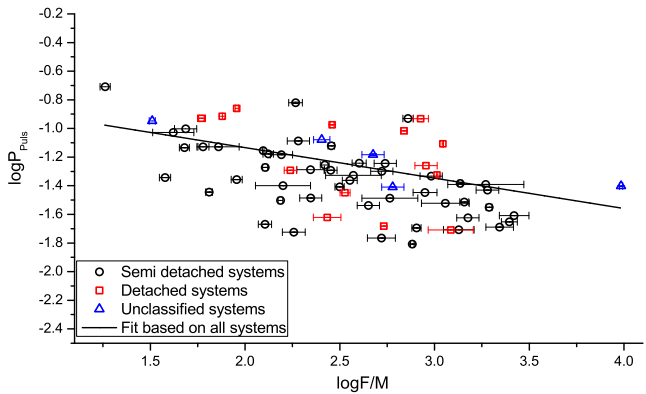}
\caption{The correlation between $\log F/M_{\rm pul}-\log P_{\rm pul}$ of binary $\delta$~Sct stars in systems with $P_{\rm orb}<13$~days.}
    \label{fig:corFP}
\end{figure}

The gravitational force per gram matter of the pulsator's surface $F/M_{\rm pul}$ expresses the influence of the companion's gravity to the $P_{\rm pul}$ of the $\delta$~Sct star. The mass transfer is not taken into account in this correlation, although it probably plays a crucial role in the $P_{\rm pul}$ modulation. However, so far, it is the only way to quantify the influence of the binarity on the dominant oscillation frequency. For the calculation of the $F/M_{\rm pul}$ the formalism of \citet[][eq.~10]{CAK16} was used. The $F/M_{\rm pul}$ is derived for each system in cgs units (i.e. dyn~gr$^{-1}$), while the $P_{\rm pul}$ is expressed in days. For the $F/M_{\rm pul}$ values for which no errors exist in the literature we assigned errors following the same method as in the previous correlation (eq.~\ref{cor:gP}). We found 0.003 as an average $F/M_{\rm pul}$ error value for SB2+E, 0.008 for SB2, 0.059 for E, 0.040 for SB2+EV, and 0.007 for SB1+E systems. The sample for the correlation between $F/M_{\rm pul}-P_{\rm pul}$ consists of 68 pulsators. In particular, 49 of them belong to semi detached, 14 to detached, and 5 to unclassified systems. In Fig.~\ref{fig:corFP} the fit on the $\log P_{\rm pul}-\log \frac{F}{M_{\rm pul}}$ values is shown. The relation between gravitational force per surface gram matter is the following:

\begin{equation}
\log P_{\rm pul}= -0.71(2)-0.213(7)~\log \frac{F}{M_{\rm pul}},~with~r=0.49
    \label{cor:FP}
\end{equation}

\section{Discussion and Conclusions}
\label{sec:s5}

The current study presents the most complete up-to-date catalogue of $\delta$~Sct stars in binaries. 203 $\delta$~Sct stars in 199 binary systems are included and they are distinguished according to the available information and their characteristics (i.e. $P_{\rm orb}$ and Roche geometry configuration). Moreover, demographics for these systems relevant to our knowledge about them, their distribution within the stellar evolutionary diagrams as well as updated correlations between their fundamental properties are also presented.

The sample of this study is divided into two groups according to the orbital period values of the systems, with the limit to be set on $P_{\rm orb}=13$~days. This limit should be considered as a non-strict one and its assumption was based on the distribution of the systems within the $\log P_{\rm pul}-\log P_{\rm orb}$ plot (Fig.~\ref{fig:statPP}). For systems with $P_{\rm orb}$ values greater than 13~days it seems that the $P_{\rm pul}$ is not affected by the binarity at all, while those having $P_{\rm orb}<$13~days form a `band' with a clear linear correlation between orbital and pulsation periods. However, there is no doubt that this arbitrary limit might move to a larger $P_{\rm orb}$ value, if more systems with $P_{\rm orb}\sim13$~days will be discovered. Thus, for the time being, in absence of more such systems, this limit seems to differentiate in a satisfactory way the binary $\delta$~Sct stars, whose dominant oscillation frequency is influenced by the presence of a companion.

The $P_{\rm pul}-P_{\rm orb}$ dependence in systems with $P_{\rm orb}<13$~days as well as the limit of $P_{\rm orb}=13$~days cannot be explained so far with high certainty and with a quantitative way. Speaking qualitatively, we can say that in short-period (i.e. $P_{\rm orb}<13$~days) detached systems the tidal effects may play an important role to the pulsation frequency modulation. Although tidal potential excites g-mode pulsations \citep{ZAH08, ZAH13, HAM16}, there is probably an influence to the p-modes and/or to the radial mode of the pulsating member of the system. In oEA stars this dependence seems to be more complicated because the mass transfer rate may be also a determining parameter additionally to the tidal interaction influence in the pulsations likely in detached systems.

Updated correlations between $P_{\rm pul}-P_{\rm orb}$ according to the geometrical status of the systems with $P_{\rm orb}<13$~days are also derived. The comparison of equations~\ref{cor:logPPSD}, \ref{cor:logPPD}, and \ref{cor:logPPall} with the respective ones of \citet[][eq.~1-3]{LIA12} shows that the correlations have not changed significantly, although the present sample is $\sim50\%$ larger than that of 2012. On the other hand, if we compare the decimal forms of $P_{\rm pul}-P_{\rm orb}$ correlations (eq.~\ref{cor:PPSD}, \ref{cor:PPD}, and \ref{cor:PPall}) with the respective one of \citet[][eq.~1]{SOY366}, we can see that the slope has been changed to almost half of its initial value. The present correlation has to be considered as more accurate since the sample is $\sim5$ times larger than that of \citet{SOY366}. Finally, since these empirical correlations show that the coefficient of $\log P_{\rm orb}$ in equations~\ref{cor:logPPSD}, \ref{cor:logPPD}, and \ref{cor:logPPall} is far away from 1, regardless of the Roche geometry configuration of the systems, the theoretical approximation made by \citet[][eq.~7]{ZHA13} may need further examination.

Figs~\ref{fig:statMR}, \ref{fig:statHR}, and \ref{fig:statgT} show the distributions of the binary $\delta$~Sct stars within the well known $H-R$, $M-R$, and $T_{\rm eff}-\log g$ diagrams. From these diagrams, nine pulsators have been found to be slightly evolved lying beyond the TAMS line. On the other hand, four $\delta$~Sct stars which are the primary components of RZ~Cas, AS~Eri, AO~Ser, and VV~UMa lie below (but very close to) the ZAMS lines of $M-R$ and $T_{\rm eff}-\log g$ diagrams, but above the ZAMS in the $H-R$ diagram. Therefore, it can be plausibly concluded that, since more than $\sim85\%$ of these stars lie between ZAMS-TAMS limits, the $\delta$~Sct stars in close binaries can be considered as Main-Sequence pulsators. Moreover, in Figs.~\ref{fig:statHR} and \ref{fig:statgT} it is shown that $\sim15\%$ of these stars lie beyond the blue edge of the instability strip, while only $\sim6\%$ of them are located below the red edge. Small discrepancies regarding the location of the stars close to the boundaries of the Instability Strip probably come from the different theoretical evolutionary models used for the ISB-ISR lines of the $H-R$ and $T_{\rm eff}-\log g$ diagrams.

From the present sample it is found that the frequencies range of the currently known $\delta$~Sct stars in binaries is 4-64 cycle~d$^{-1}$. The slowest pulsator ($f_{\rm dom}=4$~cycle~d$^{-1}$) belongs to $\gamma$~Boo, while the fastest one ($f_{\rm dom}=64.4$~cycle~d$^{-1}$) to KIC~11175495. Unfortunately, the absolute parameters of these two systems have not been calculated to date. However, based on those located outside the Main Sequence limits and mentioned in the previous paragraph, it is found that the less evolved stars pulsate with $f_{\rm dom}$ values greater than 48~cycle~d$^{-1}$, except for that of AO~Ser. On the other hand, the more evolved pulsators oscillate with $f_{\rm dom}$ values less than 20~cycle~d$^{-1}$.

The updated correlation between $\log g -\log P_{\rm pul}$ (eq.~\ref{cor:gP}) shows significant change from that derived by \citet[][eq.~5]{LIA12}. In the present study, the sample used for this correlation is $\sim35\%$ larger than that of 2012, and, in addition, a new approach regarding their reliability (i.e. errors) was used. For these two reasons, the present correlation can be considered as more realistic and better approximated. Binary $\delta$~Sct stars with $\log g>3.96$ and $\log P_{\rm pul}<-1.36$ (=$f_{\rm dom}>23$~cycle~d$^{-1}$), i.e. the younger and faster pulsators, follow very well the $P_{\rm pul}-\log g$ trend (Fig.~\ref{fig:corgP}), while the scatter around the fitted line is larger for the older and slower pulsators, especially for those belonging to semi-detached systems. This is probably related to the mass transfer occurrence between the binary members as well as to their evolution. The pulsations in binary $\delta$~Sct stars can potentially begin during the very early stages of their MS life or even before it. According to the distribution shown in Fig.~\ref{fig:stat3D}, it can be concluded that as the pulsator evolves its dominant pulsation frequency is decreased following the well known frequency evolution of the single pulsating stars. However, at the same time its companion is also evolving, although with slower rate (it should be noted that the pulsators almost always have greater mass than their companions--see Table~\ref{tab:tab1}), thus the mass transfer rate is accelerated and the pulsations are certainly affected. Speaking qualitatively, it can be said that probably there is a critical mass gain value of the pulsator after which the correlation between its evolutionary status and its dominant pulsation frequency is agitated. All the above stand for the $\delta$~Sct stars in semi-detached systems of our sample. On the other hand, the pulsators of the detached systems show slightly different behaviour, but it should be noted that the present sample of $\delta$~Sct stars in detached systems is much less than the aforementioned one. However, there are only three cases (CoRoT~105906206, HD~172189, and FL~Ori) clearly deviating from the $\log g -\log P_{\rm pul}$ fit (between $4.4<\log g<3.5$ and $-1.3<\log P_{\rm pul}<-0.96$), while the three last ones follow it very well. A careful examination of the deviating systems was made, but no obvious similarities (e.g. close mass ratio values) were found. Therefore, with the present sample, the reason of their deviation cannot be explained and more systems have to be added for more certain conclusions.

In order to compare the present correlation between $\log \frac{F}{M_{\rm pul}}-\log P_{\rm pul}$ with that derived by \citet[][eq.~7]{SOY13DRA} a change in the $P_{\rm pul}$ measurement units of equation~\ref{cor:FP} is needed. Therefore, by expressing the $P_{\rm pul}$ in sec instead of days, the intercept of equation~\ref{cor:FP} becomes 4.2(6). Thus, we can see that the intercepts of the compared equations are quite similar, if we take into account the error bars, but, contrary to that, their slopes differ significantly. We found a slope of -0.213(7), while the respective value of \citet{SOY13DRA} is -0.40(6). At this point, it has to be noted that the approximations for this correlation are also different between these two studies. We used a sample of 68 stars by taking into account the reliability of their absolute parameters (i.e. errors), while \citet{SOY13DRA} were based only on the 21 cases with accurate measured absolute parameters. If we use only the stars of our sample with the highest certainty on their absolute parameters (i.e. those that belong to SB2+E systems), then a relation very similar to that of \citet{SOY13DRA} is derived.

The geometrical configurations of the unclassified eclipsing systems are expected to be determined after a simple light curves analysis. For the ellipsoidal variables and for the spectroscopic binaries, which might also be eclipsing ones but they have not been observed yet, maybe photometric modelling is feasible. Contrary to these, for the spectroscopic non-eclipsing binaries, for the systems whose binarity was discovered by the $O-C$ variation of their $f_{\rm dom}$, and for the visual binaries we cannot expect any Roche geometry determination.

The most important thing for the future study of this group of stars is undoubtedly the enrichment of the current sample in order to establish further and with higher certainty the present correlations. However, in order to determine more accurately the absolute parameters of the currently known systems, spectroscopic surveys, especially for the eclipsing ones, are strongly recommended, given that the combination of RVs and light curves modellings provides the ultimate knowledge for a system. Theoretical approximations regarding the connection between $P_{\rm pul}-P_{\rm orb}$ as well as the limit of 13 days in $P_{\rm orb}$ are also of high importance, since only a one such study exists so far. Regarding new discoveries, good targets for ground-based photometry and/or spectroscopy can be considered the cases whose pulsations type are unclear (see Table~\ref{tab:tab4}). In addition, new observational campaigns and extensive search in the data of space missions such as $Kepler$ are needed to be done in order to discover new cases.

\section*{Acknowledgments}
A.L. acknowledges financial support by the European Space Agency (ESA) under the NELIOTA programme, contract No.~4000112943. This research has made use of NASA's Astrophysics Data System Bibliographic Services and the SIMBAD database, operated at CDS, Strasbourg, France. We thank the anonymous reviewer for the valuable comments that improved the quality of the present work.

\bibliographystyle{mnras}
\bibliography{ref}

\bsp
\label{lastpage}
\end{document}